\documentclass{article}

\usepackage[english]{babel}

\usepackage{subcaption}
\usepackage{libertine} 
\usepackage{siunitx}

\usepackage[letterpaper,top=2cm,bottom=2cm,left=3cm,right=3cm,marginparwidth=1.75cm]{geometry}

\usepackage{amsmath}
\usepackage{amssymb}
\usepackage{graphicx}
\usepackage[colorlinks=true, allcolors=blue]{hyperref}

\title{The role of symmetries in the axisymmetric jet mean velocity profile development}
\author{Preben Buchhave$^1$, Mengjia Ren$^2$ and Clara Marika Velte$^3$\footnote{1. Intarsia Optics, Sønderskovvej 3, 3460 Birkerød, Denmark.
2. Department of Aeronautics and Astronautics, Kyushu University, 744 Motooka, Nishi-ku, Fukuoka 819-0395, Japan .
3. Department of Civil and Mechanical Engineering, Technical University of Denmark, Koppels Allé, Building 403, 2800 Kongens Lyngby, Denmark.}}

\begin{document}
\maketitle

\begin{abstract}
The fact that physical conservation laws can be derived from symmetry properties of space and time, as shown by Emily Nöther, has been utilized in predicting the development of the round turbulent jet from the jet exit to the far field. In particular, the developing region has been described using an analytical approach in combination with using a numerical recursive program. Both approaches assume that the only forces acting on the flow are internal shear forces in a Newtonian constant density fluid. The analytical and numerical predictions both display excellent agreement with carefully conducted measurements. The jet spreading angle is observed to be directly coupled to the turbulent momentum diffusion, hence the spreading rate depends on the upstream, or initial, conditions. The jet entrainment and momentum rate are both observed to be constant, even across the developing jet. Since the solution of the jet development depends on the initial conditions, the total (molecular and turbulent) viscosity and the initial velocity profile must be input into the analytical or numerical solver to obtain the correct solution. The Reynolds number is observed to not enter into the analytical or numerical solution and experiments confirm that the jet spreading is independent of the Reynolds number in the tested range, $Re = 3\,200 - 32\,000$. We emphasize that our results do not rely on any assumptions of self-similarity or prior knowledge about the jet from experiments, only the Galilei symmetry properties, and that the results are valid throughout the jet. 
\end{abstract}

\section{Introduction}

``Free flows'' are not really free, but subject to the conditions imposed upon them by the so-called symmetry properties of time and space. In most normal environments, these properties are assembled in the Galilei Group as space translation symmetry, time translation symmetry, rotational symmetry about a chosen axis and reference frame translation symmetry. According to Nöther’s theorem~\cite{noether1983invariante}, each of these symmetry groups can be expressed mathematically by a physical law. All phenomena, in particular fluid flows, must adhere to these laws.

The role of symmetry in nature and physics has long been recognized and is described in for example~\cite{schwichtenberg2018physics}. Symmetry in fluid flows is described in e.g.~\cite{frisch1995turbulence}. Symmetry is also applied in mathematics, where symmetry in, for example, the governing differential equations in fluid flow can be reduced in complexity by utilizing symmetry, see for example~\cite{oberlack1999symmetries,oberlack2000symmetrie,oberlack2002turbulence}. 

We have studied a well-known canonical flow phenomenon, the free, axisymmetric jet in air, and described what basic properties can be predicted for this flow based on the Galilei symmetries and fundamental physical assumptions. In earlier publications,~\cite{buchhave2022similarity,zhu2022similarity}, we studied the jet at a distance far from the jet exit, the so-called self-similar region, and showed that self-similarity is really an unavoidable consequence of the Galilei symmetries. We could predict that the jet, in the self-similar region, must in the average spread as a linear cone with its top point in a virtual center close to the jet exit. We also showed from symmetry considerations and verified by laser Doppler measurements, that all statistical properties can be scaled by a single factor, namely the axial distance to the virtual center.

In the present work, we present results based on symmetry properties and basic physical assumptions, related to the mean velocity profile as it develops from the initial profile at the exit and until it has developed into the self-similar state. The study also allows us to predict the jet width and spreading angle as well as the mass entrainment in both the developing region and the self-similar region.	

Nonequilibrium developing turbulence is a field of fluid mechanics of great interest for the understanding of turbulent flows and for practical applications and turbulence modelling. The free, axisymmetric jet in air is an ideal research item for the study of nonequilibrium turbulence since the flow after leaving the environment of the jet exit is unaffected by external forces and left to its own devices to develop from an initial incident state to a final self-similar state. Self-similarity for the axisymmetric jet is, of course, well documented, see for example~\cite{monin1971vol,Pope_2000}. See also~\cite{hussein1994velocity,burattini2005similarity} for large experimental studies of the axisymmetric jet in the self-similar region using X-hotwires and~\cite{george1989self} for self-preservation of homogeneous turbulent shear flows. However, the description of self-similarity is usually based on assumptions, later to be verified by comparison to experimental results or by demonstrating accordance with the basic equations of motion. In contrast, in~\cite{buchhave2022similarity} and in the present work, we have derived the properties of the jet from an arbitrary incident state at the jet exit to a final self-similar state by using only the fundamental symmetry properties of the Galilean Symmetry Group and by using the fact that air is a Newtonian constant density fluid. Careful measurements, recorded in~\cite{zhu2022similarity}, confirm the one-factor scaling of statistical quantities. 

Using the symmetry properties, we have developed a small, recursive computer program that can run on a single regular PC, that computes the radial profile of the mean of the jet velocity at any downstream distance from the jet orifice and out through the self-similar region. From these computations, we can see that the axisymmetric jet has the property that any arbitrary initial velocity profile for a given jet geometry will develop into a self-similar velocity profile downstream at distances beyond 20-30 jet exit diameters. The final width and spreading angle of the mean velocity profile will depend both on the turbulent viscosity (averaged over radial distance) and on specific geometric properties of the orifice. These results are computed and plotted in a cylindrical coordinate system, $(r,\varphi,z)$, with the $z$-axis along the jet symmetry axis.

Based on the recursive computer algorithm, we have derived a second-order partial differential equation in the radial coordinate $r$ and time $t$. This equation is equivalent to the heat equation in cylindrical coordinates and describes the momentum diffusion caused by the turbulent kinematic viscosity. We have solved this equation by the Lie method, see for example~\cite{güngör2024notesliesymmetrygroup} or~\cite{oberlack2002turbulence}, and obtained an analytical expression for the velocity profile. The solution matches very closely the results of the recursive algorithm. Standard textbooks, e.g.~\cite{Pope_2000} and earlier derivations by~\cite{schlichting1933laminare}, use a boundary layer approximation to the jet shear layer which results in an analytical solution for the mean velocity profile in the self-similar region. However, we emphasize that our derivation is based on basic symmetries and physical properties of the flow, whereas the quoted texts assume self-similarity and use the basic equations of motion in some simplified form to arrive at an equation for the self-similar velocity profile. Below we shall discuss the differences between our analytical solution and the results of our computations and measurements.

We have compared the theoretical mean velocity profiles at different distances along the jet axis to experiments performed on three different jets with exit diameters of 100 mm, 50 mm, and 10 mm. The measurements were performed by hotwire anemometry (HWA) and laser Doppler anemometry (LDA), and the results demonstrate excellent agreement between the measurements and the theoretical predictions.

In the following, we shall also display other basic jet properties based on the numerical difference equation. For example, we find that without explicitly having assumed momentum conservation at any point in the derivation, the computation shows a constant rate of momentum transport all the way from the exit to beyond the self-similar region. We also display the development of the average centerline velocity both as a function of time and as a function of axial distance. 

We notice slight differences in spreading angle for different jet orifice geometries and we demonstrate the dependence of the jet properties on the turbulent structures reflected in the turbulent kinematic viscosity (eddy viscosity) even in the far field. Inclusion of these subtleties in the computer program requires the mean velocity to have a small component in the $r$-direction. We compute the conservation of rate of momentum transport as well as plots of the jet width and mass entrainment as a function of axial distance, demonstrating the influence of the initial properties and the turbulent structures on the properties of the jet in the far field.

\section{Computer program -- derivation and algorithm}

\begin{figure}
\centering
\includegraphics[width=0.75\linewidth]{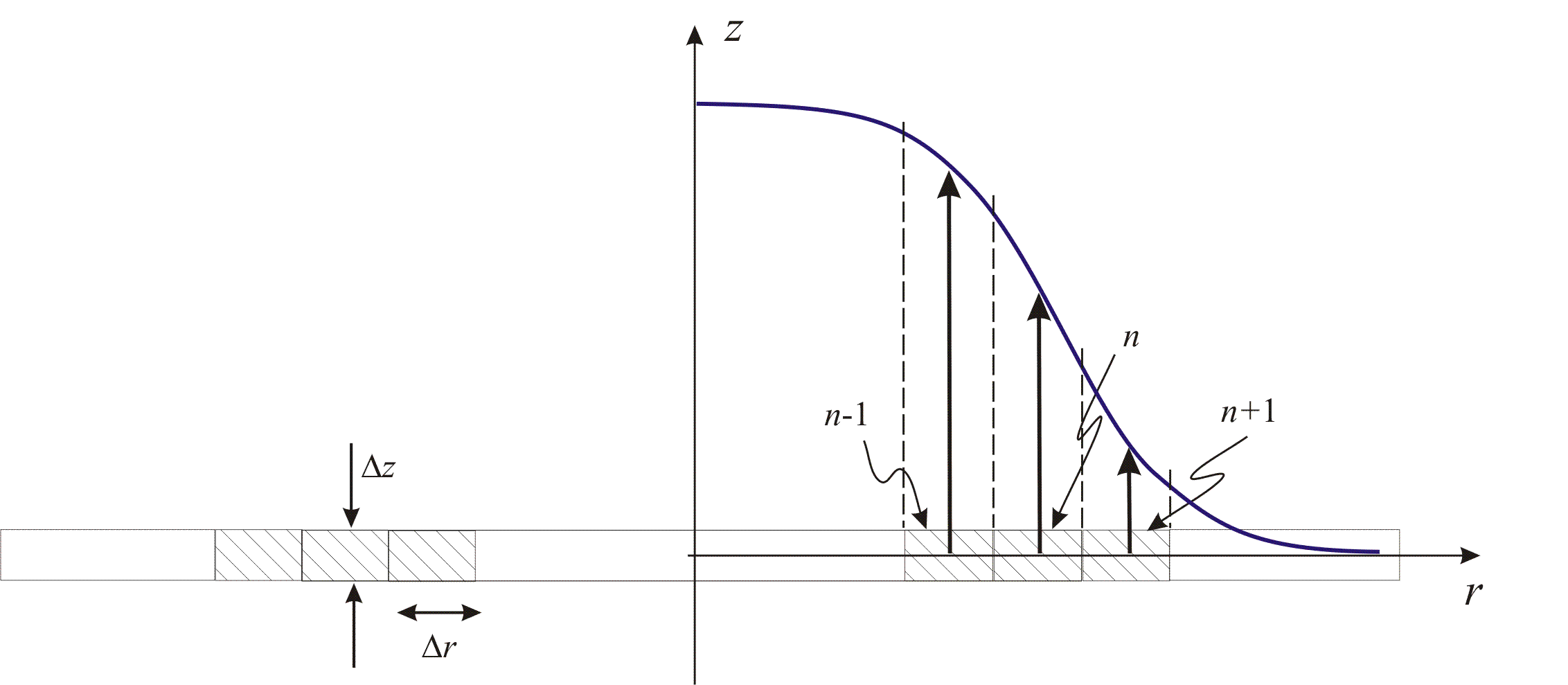}
\caption{\label{fig:numericalsetup}Disc-shaped control volume with associated ring-elements from which the mean velocity profile is derived.}
\end{figure}

The model is based on two basic properties of the stationary flow: 
\begin{enumerate}
    \item The flow is circularly symmetric around a direction in space, the jet axis.
    \item The only forces acting on the flow are internal shear forces acting in a Newtonian constant density fluid.
\end{enumerate} 
We have thus neglected dynamic pressure, which for subsonic flows is small compared to the shear forces, and dissipation, which means we limit our consideration to the high Reynolds number inertial range, which is typical of many medium to high Reynolds number axisymmetric jet flows. Due to the axial symmetry, the problem is reduced to two dimensions in a cylindrical coordinate system, namely the jet axial direction $z$ and the radial direction $r$, see Figure~\ref{fig:numericalsetup}. 

In our model, averaging turbulent fluctuations, we consider the time development of the mean transport of momentum through a small ring concentric with the $z$-axis as depicted in Figure~\ref{fig:numericalsetup} in the cylindrical coordinate system aligned with the flow, i.e. with the $z$-axis along the jet axis, the radial coordinate $r$ and the azimuthal angle $\varphi$. At the downstream position $z$, we consider a thin disc control volume consisting of a number of concentric ring elements. Figure~\ref{fig:numericalsetup} shows three consecutive ring-shaped fluid elements of thickness $\Delta z$ in the axial direction and of width $\Delta r$ in the radial direction. The rings are numbered $n-1$, $n$, $n+1$ at the radial positions $r_{n-1}$, $r_n$ and $r_{n+1}$ where $n$ is a radial index. The mean velocity in the axial direction of the $n$'th ring element is denoted $u_n$. We need the streamwise velocity gradient in the radial direction of the inner and the outer surface of the $n$’th ring, $G_{n,i}$ and $G_{n,o}$, and the inner and outer areas of the $n$’th ring, $A_{n,i}$ and $A_{n,o}$, to compute the shear forces on the surfaces of the ring and the resulting momentum change. To compute the velocity from the momentum, we also need the mass of the $n$'th ring, $M_n = \rho V_n$, where $Q$ is the density of air and $V_n$ is the volume of the $n$'th ring.

We proceed by assuming a known initial mean velocity of all the rings, $u_{in}$. We then compute the shear forces on the inner surface and the outer surface of the $n$'th ring, $f_{n,i}$ and $f_{n,o}$. The total force on the ring element is then $f_n = f_{n,i} - f_{n,o}$. The shear forces are given by $f_{n,i} = \rho \nu_{Total} G_{n,i} A_{n,i}$ and $f_{n,o} = \rho \nu_{Total} G_{n,o} A_{n,o}$, where $\nu_{Total}$ is the general total kinematic viscosity $\nu_{Total} = \nu + \nu_T$, i.e., including both molecular viscosity, $\nu$, and (radially averaged) turbulent eddy viscosity, $\nu_T$. We estimate the gradient by using the velocity on either side of the surface, $G_{n,i} = (u_n-u_{n-1})/\Delta r$ and $G_{n,o} = (u_{n+1}-u_{n})/\Delta r$. We then compute the incremental change in velocity from the change in momentum for the ring, $\rho V_n u_n$, in a small time-increment, $\Delta t$: 
\begin{equation}
    \Delta u_n = \frac{f_n \Delta t}{\rho V_n}
\end{equation}
By repeatedly computing the incremental changes to the velocity, we can see the change in the shape of the velocity profile as a function of time.

The calculation must be initiated with an incident velocity profile. This can be a numerical array derived from an analytical expression or it can be measured velocities. The program then repeats the calculation of incremental changes to the momentum through each ring during the small time-increment, $\Delta t$. We have chosen as a first example of an analytical incident profile, $u_0(r)$, a so-called super Gauss with center value $U_0$,
\begin{equation}
    u_0(r)=U_0e^{-\frac{1}{2}\left ( \frac{r}{R} \right )^p}
\end{equation}
which is the ordinary Gaussian function elevated to the power $p$, where $R$ is the radius of the jet orifice. The ``super Gauss'' profile is close to the often assumed top-hat profile at the exit of many laboratory jets. Later we shall use measured mean velocity profiles near the jet exit as input to the program.

\section{Differential equation}

In the limit of an infinitely thin disk and with infinitely small time steps, the difference equation becomes a two-dimensional partial differential equation for the time development of the mean velocity:
\begin{equation}\label{eq:PDEcyl}
    \frac{\partial u(r,t)}{\partial t} = \nu_{Total} \left ( \frac{1}{r} \frac{\partial u(r,t)}{\partial r} - \frac{\partial^2 u(r,t)}{\partial r^2} \right )
\end{equation}
This equation for the diffusion of momentum is similar to the heat conduction equation, which describes the diffusion of heat.  We expect a solution in terms of error functions as one obtains for the heat conduction equation. It resembles the ordinary differential equation applied by~\cite{Pope_2000} and~\cite{schlichting1933laminare}, but our equation describes the time development of the momentum of a thin disk normal to the jet axis based on the momentum diffusion due to internal shear forces, while the equation of~\cite{Pope_2000} is based on a simplified boundary layer equation and is valid only for the self-similar region. Our equation describes the developing turbulent flow as well as the flow in the self-similar region. This will be verified in the following through comparisons between measurements and computations. 

\cite{Pope_2000} arrives at an analytical solution to the boundary layer equation,
\begin{equation}
    u(\eta) = U_0 \frac{1}{ \left ( 1+a\eta^2 \right )^2}
\end{equation}
where $\eta = \frac{r}{(z-z_0)}$ is a self-similar radial coordinate scaled with the distance from the self-similar virtual origin $z_0$, $U_0$ is the center velocity at the exit and $a$ is a constant that must be determined experimentally.  This solution applies only to the self-similar flow beyond 30 jet exit diameters and does not display the time development of the developing jet. In addition, the solution of~\cite{Pope_2000} extends to infinite radius in contrast to real jets that have a definite boundary to the surrounding quiescent fluid. We later compare graphs showing the solution of~\cite{Pope_2000} and our analytical and computer program solutions.

To solve the second order partial differential equation (\ref{eq:PDEcyl}), we apply the symmetries of the equation and use the Lie method~\cite{güngör2024notesliesymmetrygroup} to convert the second order partial differential equation to an ordinary second order differential equation, which we integrate. Assuming an initial velocity profile, $u_0(r,0)$, with center velocity $U_0$, our analytical solution is given as the convolution of the initial profile $u_0(r,0)$, 
\begin{equation}
    u(r,t) = u_0(r,0) \circledast U(r,t)
\end{equation}
with a kernel, 
\begin{equation}
    U(r,t) = \frac{u_0}{2t}\exp \left ( -\frac{r^2}{4\nu_{Total} t} \right )
\end{equation}
that constitutes the solution to the differential equation (\ref{eq:PDEcyl}).

\begin{figure}
\centering
\includegraphics[width=0.65\linewidth]{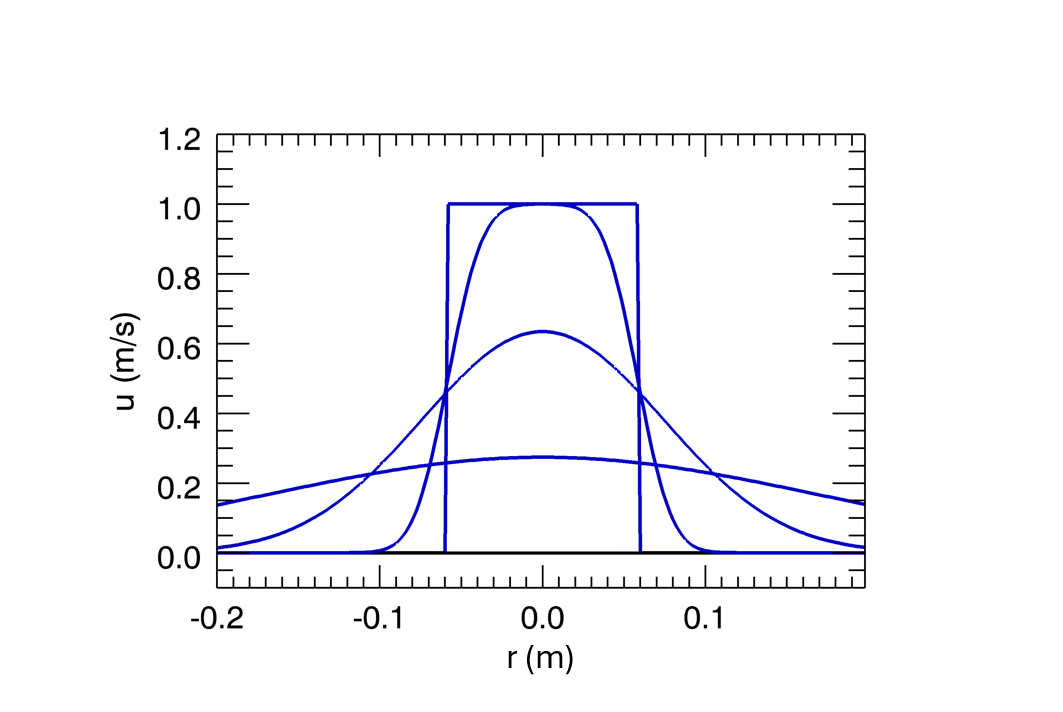}
\caption{\label{fig:veldistrasafcnoftime}The error function solution, equation (\ref{eq:erfsoln}), for increasing values of time: 0, 400, 1000, 10 000 s.}
\end{figure}

The final velocity distribution is thus obtained as the convolution of the kernel and the top-hat initial velocity distribution:
\begin{equation}
    u(r,t)=\frac{U_0}{2}\int_{-1/2}^{1/2}\frac{1}{t} \exp \left ( -\frac{(r-r')^2}{4 \nu_{Total} t} \right )\, dr'
\end{equation}
Integrating the above, we find the velocity distribution as a function of time, see Figure~\ref{fig:veldistrasafcnoftime}, showing development from the initial top-hat profile to increasingly broader velocity distributions:
\begin{equation}\label{eq:erfsoln}
    u(r,t) = \frac{U_0}{2}\left [ \mathrm{erf}\left ( \frac{r+1/2}{\sqrt{4 \nu_{Total} t}} \right ) - \mathrm{erf}\left ( \frac{r-1/2}{\sqrt{4 \nu_{Total} t}} \right )\right ]
\end{equation}

\begin{figure}
\centering
\includegraphics[width=0.50\linewidth]{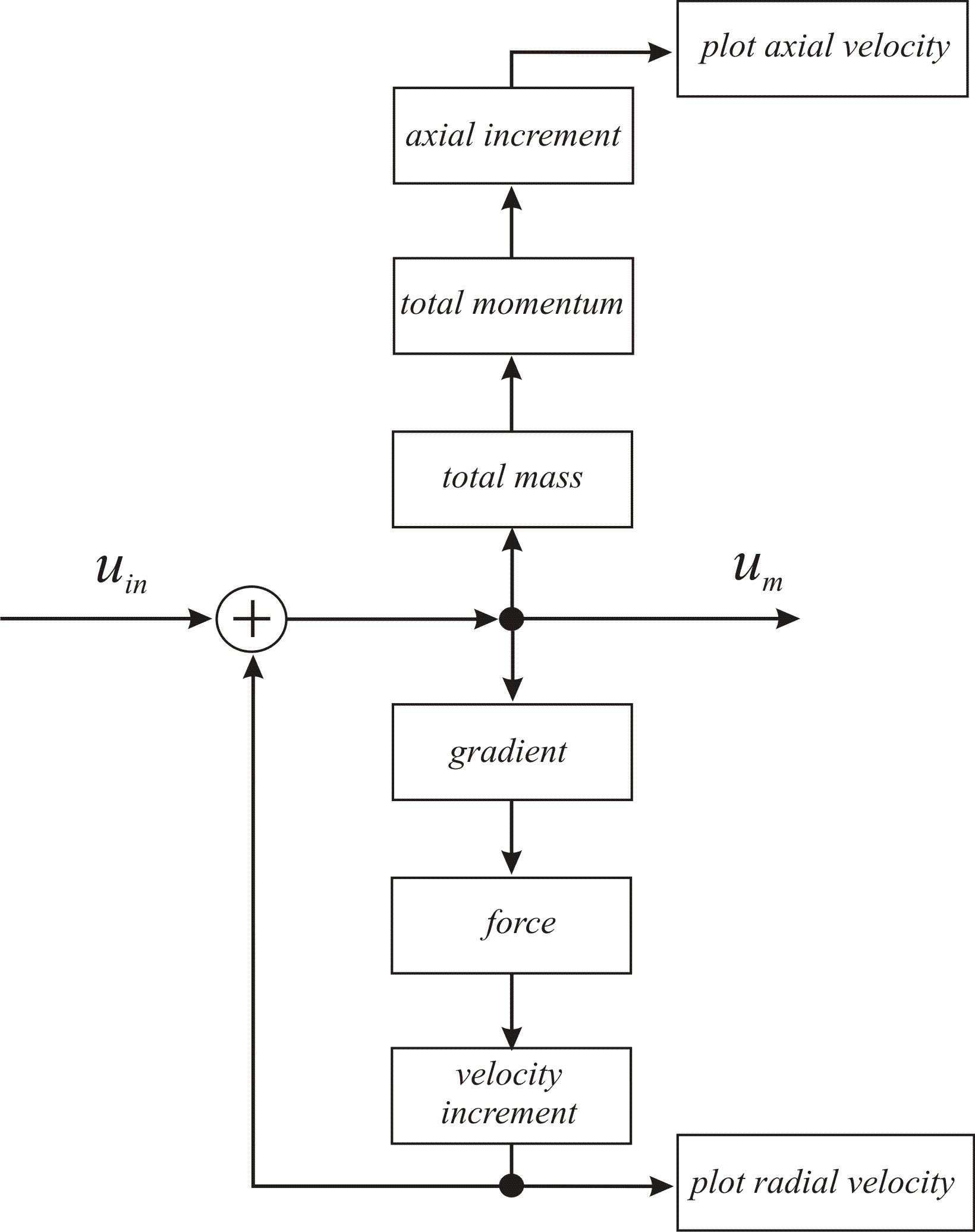}
\caption{\label{fig:algorithm}Block-diagram showing the recursive algorithm.}
\end{figure}


\section{Numerical results}

\subsection{Algorithm}

Figure~\ref{fig:algorithm} shows the path through the recursive algorithm. The computation starts with an assumed initial radial velocity profile for the axial velocity component, $u_0(r_n)_{in}$, where $n$ indicates the discrete radial position. The velocity profile at time $t=m \Delta t$ is denoted $u_m(r_n)$, where $m$ is a time index. Further inputs to the program are the physical constants, the radially averaged turbulent kinematic viscosity, $\nu_T$, and the density and molecular viscosity of air, $\rho$ and $\nu$. We also need to define the maximum radial distance, $r_{max} = N \Delta r$, where $\Delta r$ is the spatial sampling interval and $N$ is the number of radial elements and the maximum computing time, $t_{max} = M \Delta t$, where $\Delta t$ is the temporal resolution and $M$ is the maximum number of time steps. The algorithm computes at time $t=m\Delta t$ the incremental addition to the rate of momentum transport in the $n$'th ring, resulting from the friction forces from the neighboring rings in the time increment $\Delta t$. The incremental velocity of the ring is then determined by dividing the momentum increase with the mass of the ring, $\rho V_n$. We thus arrive at an expression for the mean velocity profile after a time, $t_{max}$. Section~\ref{sec:conversion} details how this time record can be converted to axial distance.

\subsection{Examples of numerical computations}

\begin{figure}[!h]
\centering
\includegraphics[width=0.5\linewidth]{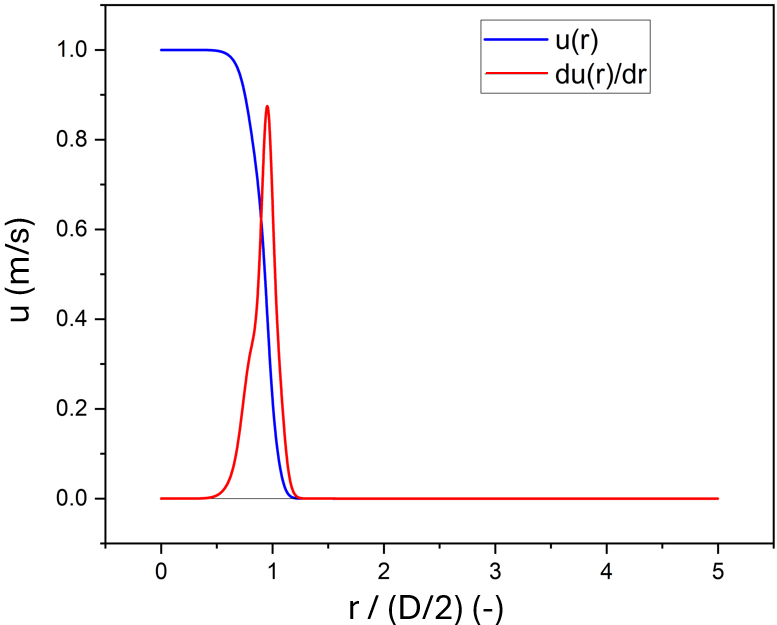}
\caption{\label{fig:GSandvelocitygradient}Initial top-hat like (super Gaussian, $p=8$) velocity profile and corresponding velocity gradient.}
\end{figure}

Figure~\ref{fig:GSandvelocitygradient} shows the initial ``super Gaussian'' profile with $p=8$ and the corresponding velocity gradient. The profile is normalized to an initial maximum velocity of unity and an initial half-width of unity.

\begin{figure*}[!h]
    \centering
    \begin{subfigure}{0.5\textwidth}
        \centering
        \includegraphics[width=\linewidth]{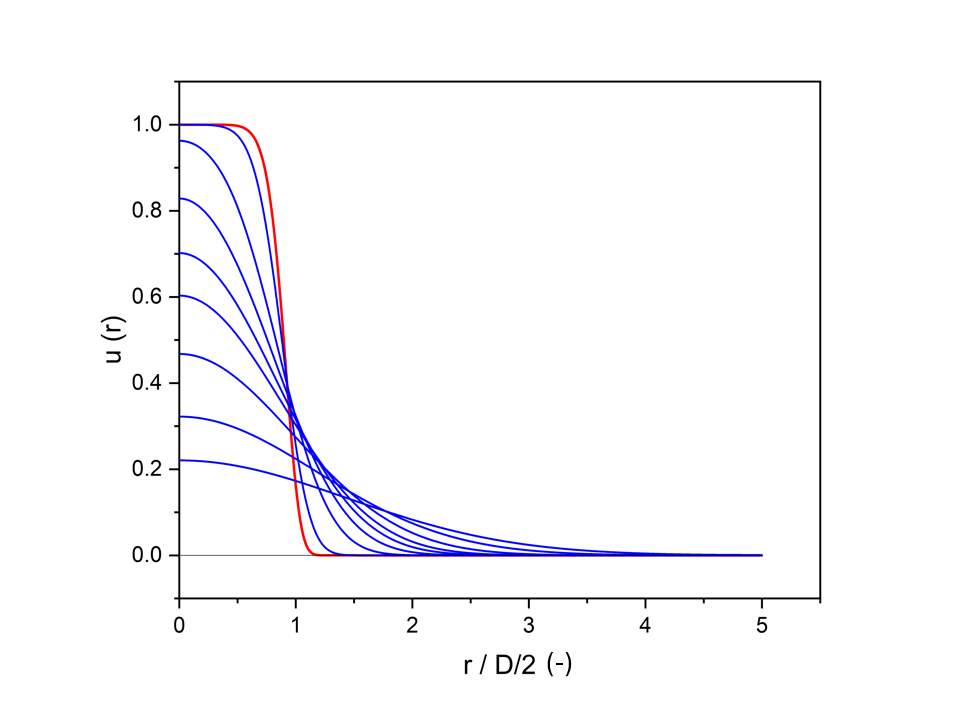}
    \end{subfigure}%
    ~ 
    \begin{subfigure}{0.5\textwidth}
        \centering
        \includegraphics[width=\linewidth]{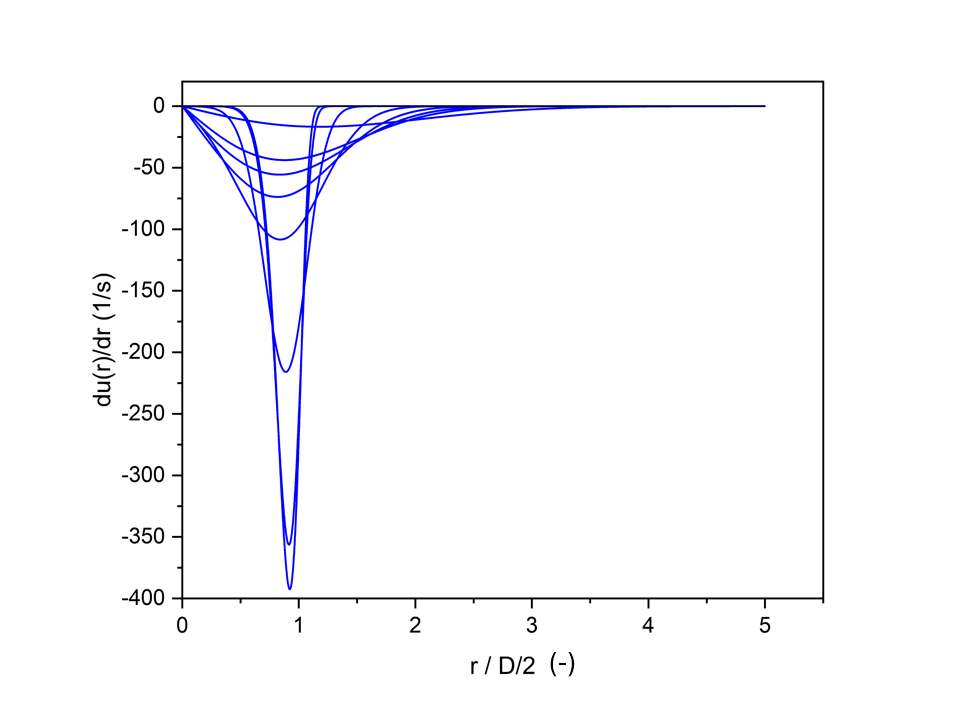}
    \end{subfigure}
    \caption{\label{fig:timedevSG} (a) Time development of the initial ``super Gauss'' velocity profile. (b) Corresponding velocity gradients.}
\end{figure*}

Figure~\ref{fig:timedevSG}(a) shows the temporal or downstream development of the ``super  Gaussian'' with an increasing number of time steps from $m=100$ to $m=80\,000$ in normalized coordinates. Figure~\ref{fig:timedevSG}(b) shows the corresponding velocity gradients.

\begin{figure}[!h]
\centering
\includegraphics[width=0.5\linewidth]{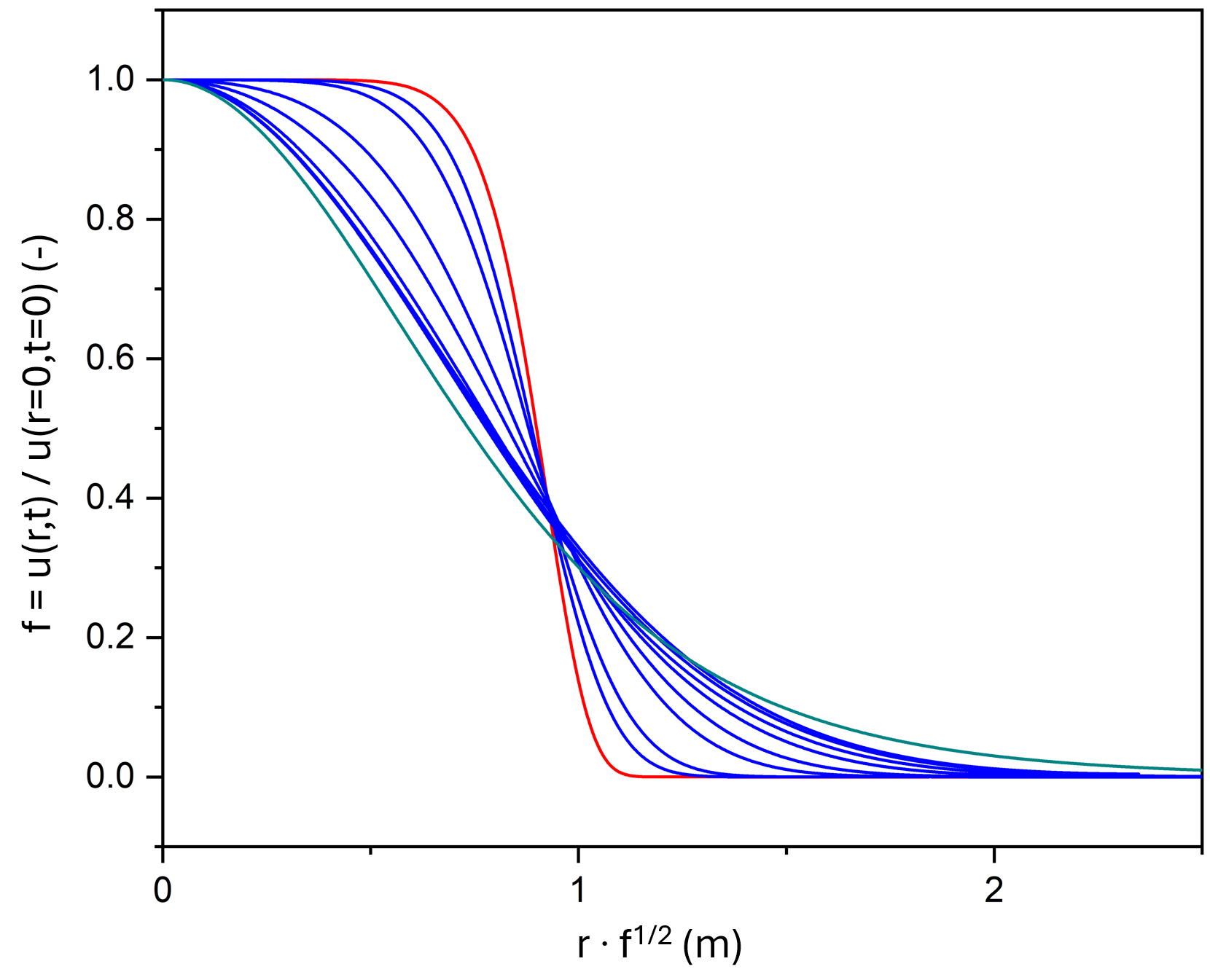}
\caption{\label{fig:scaledvelocityprofiles}Velocity profiles scaled to bring out the constant momentum flow rate between the profiles. The blue curves after 20 000 time steps collapse indicating the onset of self-similarity. The red curve is the initial velocity profile, the green curve is the scaled result of the self-similar mean velocity profile in~\cite{Pope_2000}.}
\end{figure}

Figure~\ref{fig:scaledvelocityprofiles} shows the mean velocity profiles of the flow development, scaled with the ratio between the initial center velocity and the computed center velocity in order to normalize the plot amplitude. A computation of the integrated mean momentum transport rate shows that momentum transport is indeed conserved by the calculation without momentum conservation being initially assumed; only implied by assuming Galilean symmetry.

\subsection{Self-similarity}

The blue curves in Figure~\ref{fig:scaledvelocityprofiles} begin to collapse at about $20\,000$ time steps, indicating self-similarity at longer times corresponding to greater distances indicating a single-parameter scaling. The green curve is the mean velocity profile given by~\cite{Pope_2000} also scaled to unity center velocity. The curve in~\cite{Pope_2000} is slightly different and extends to infinity whereas our calculations show a zero velocity at a finite radial distance growing with time as the jet expands.

\begin{figure*}[!h]
    \centering
    \begin{subfigure}{0.5\textwidth}
        \centering
        \includegraphics[width=\linewidth]{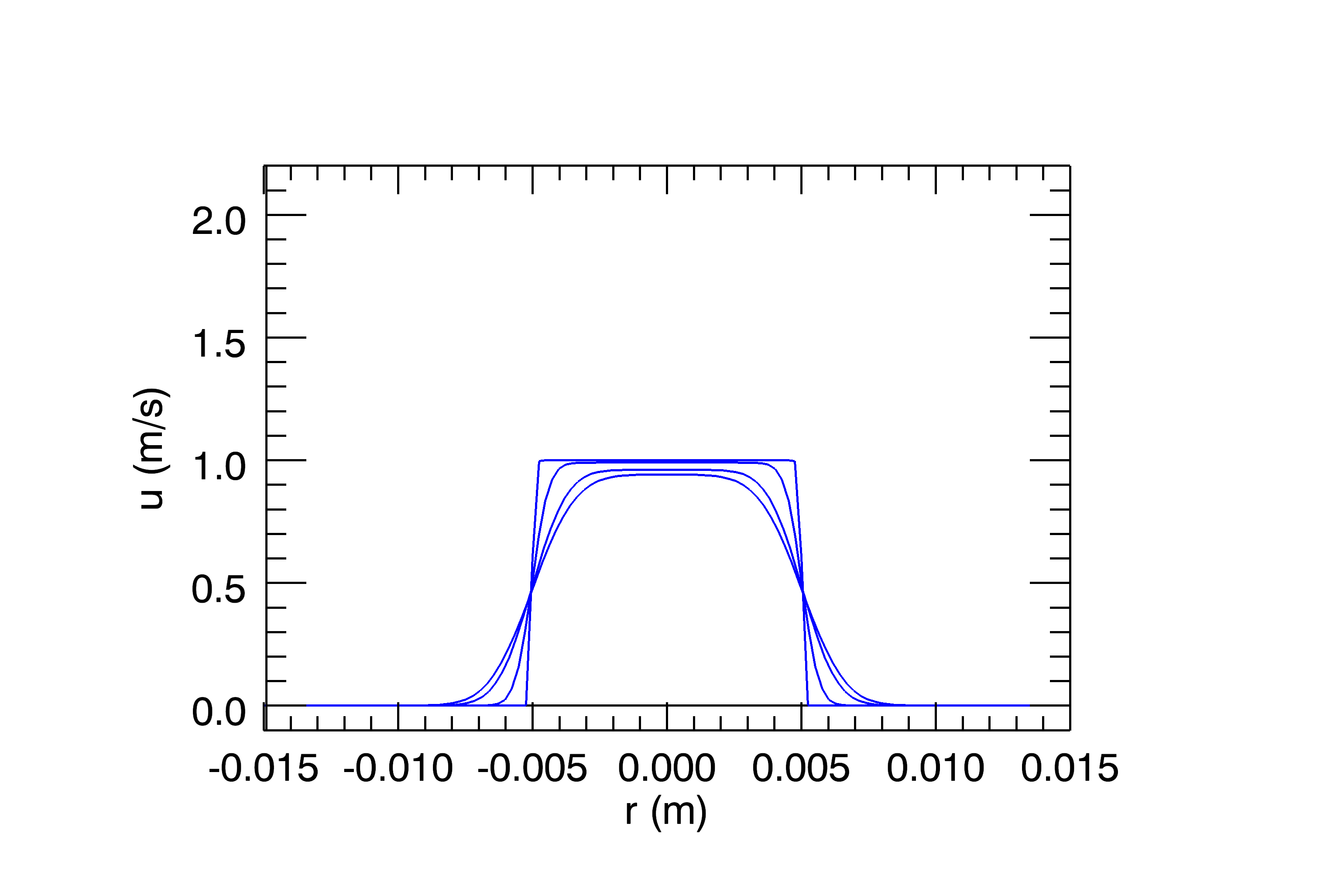}
    \end{subfigure}%
    ~ 
    \begin{subfigure}{0.5\textwidth}
        \centering
        \includegraphics[width=\linewidth]{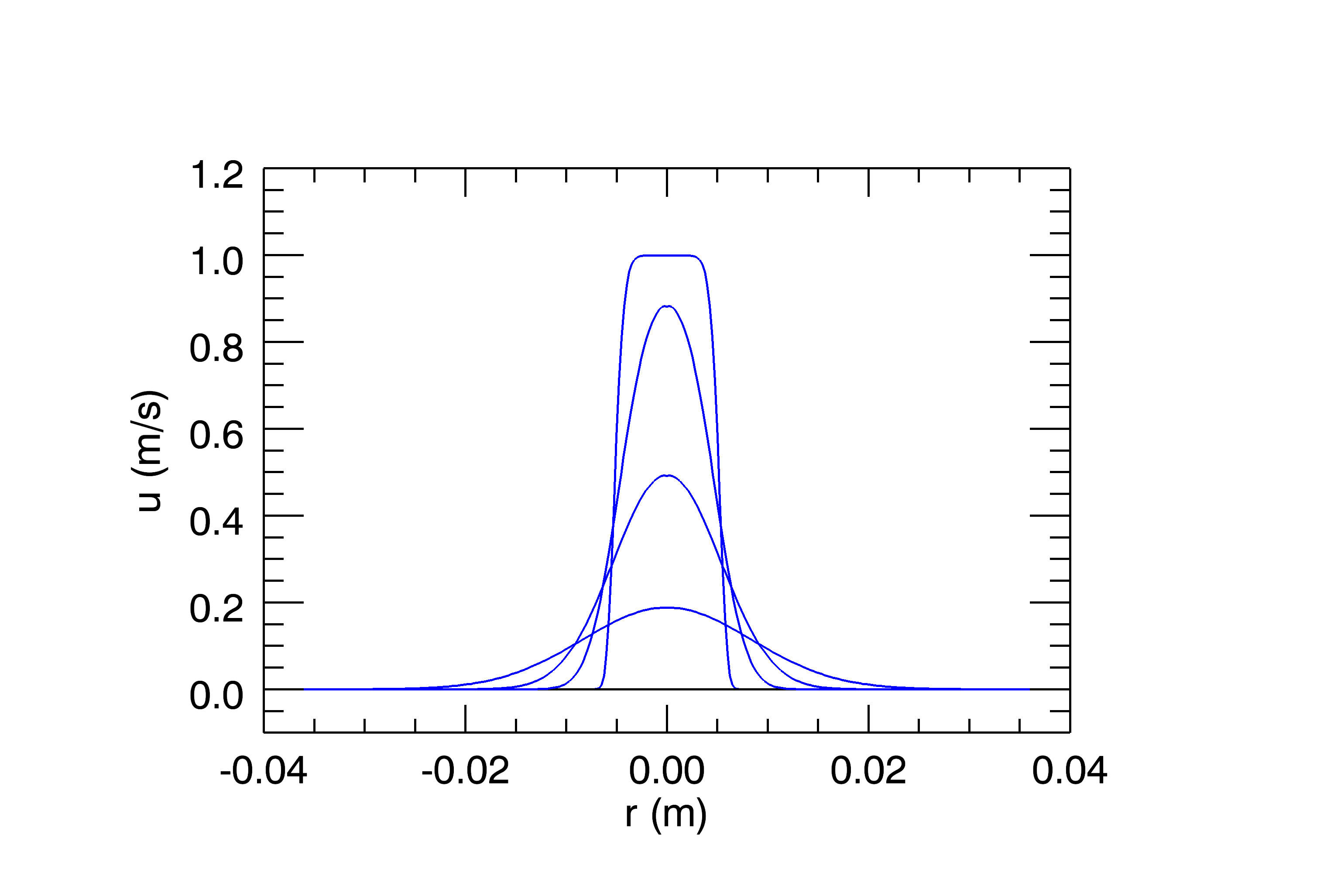}
    \end{subfigure}
    \caption{\label{fig:10mmjet} (a) Plots of computed mean velocity profiles for a 10 mm aperture jet at respectively 1, 20, 50, 100 time steps. (b) Profiles at respectively 1, 100, 1000, 10 000 time steps. $\Delta t =1.0 \cdot 10^{-4}$ s, $\nu_{Total} =1.5 \cdot 10^{-5}\, \mathrm{m^2 s^{-1}}$ (i.e., nearly laminar flow).}
\end{figure*}

Figure~\ref{fig:10mmjet} shows the development of the mean velocity profile for a 10 mm diameter jet for increasing time. Figure~\ref{fig:10mmjet}(a) displays the near field after a relatively small number of time steps. Figure~\ref{fig:10mmjet}(b) shows the combined near and far field time development with a much greater number of time steps.


\subsection{Conversion of time record to axial distance}\label{sec:conversion}

So far, we have considered the development of the mean velocity as a function of time. To compare to experiments, we need to find the development of the velocity cross section as a function of axial distance. We simply use the definition of velocity, $dz = u \, dt$, to obtain the corresponding spatial discrete distances using the time step array in the calculation. 

\begin{figure}
\centering
\includegraphics[width=0.5\linewidth]{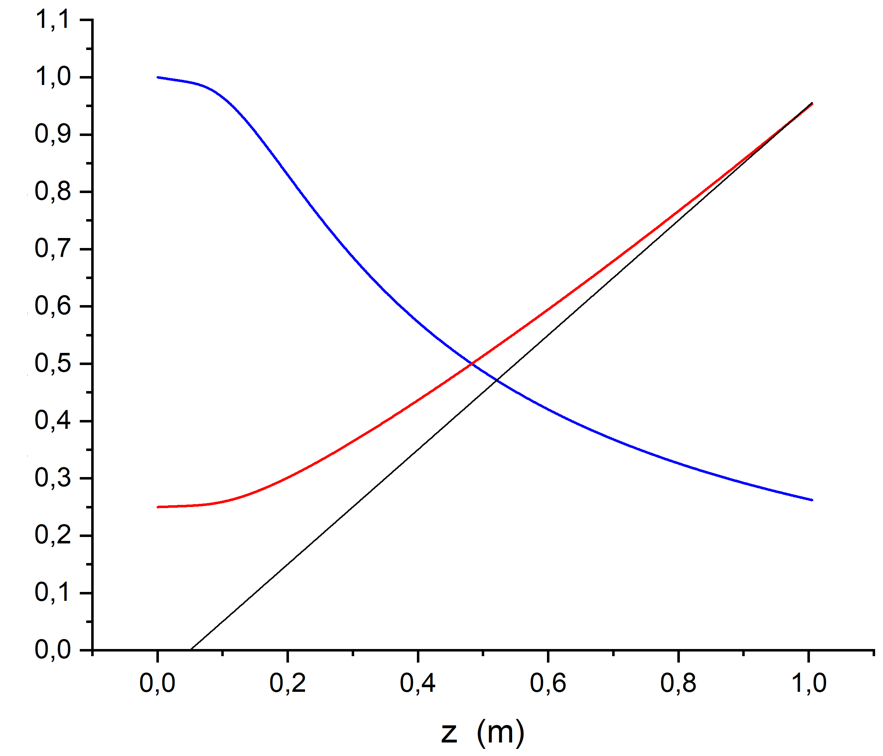}
\caption{\label{fig:centerlinevel} Development of the normalized average centerline velocity $\frac{u(r=0,t)}{u(r=0,t=0)}$ (blue curve) and the inverted average centerline velocity (red curve) for an initial super Gauss profile, with a diameter of 0.01 m, as a function of axial distance, $z$. The linear function (straight black line) indicates the self-similar velocity solution originating from $x=0.05$ m.}
\end{figure}

But what velocity should be used for the conversion? Using the center velocity results in a much overvalued distance. The velocity profile of course includes all velocity values from zero to the center velocity. We use the fact that the total momentum transport rate of the jet is conserved. We compute the total momentum, $b_m$, in a cross section at time $m \Delta t$ and integrate the mass $m_m$ over the cross section. We then form an average equivalent velocity and use that for the conversion: 
\begin{equation}
    u_{e,m} = \frac{b_m}{m_m}
\end{equation}
We can then plot the $z$-distance increment $\Delta z_m = u_{e,m} \Delta t$ and the total distance at time $m \Delta t$: $z_m = \sum_{m=0}^{m_{total}}\Delta z_m$. Figure~\ref{fig:centerlinevel} shows how the axial velocity from an analytical super Gauss with an initial cross section of 10 mm diameter develops as a function of downstream distance. The exact self-similar velocity is barely achieved even after 100 jet exit diameters downstream.

\subsection{Spreading angle, entrainment and momentum rate}

The spreading of the jet is computed by the recursive program simply by noting the radial distance for the velocity profile crossing a given level. The physics driving the jet spreading is primarily the momentum diffusion controlled by the turbulent dynamic viscosity, in addition to the significantly weaker effect from the molecular viscosity. The turbulent viscosity is not a basic physical property of the flow, but rather a quantity introduced exactly with the purpose of describing the turbulent momentum transport, so it is not surprising that we can control the computed jet properties by adjusting the turbulent viscosity. This agrees with experiments where the jet spreading angle has been modified by controlling the turbulence generation at the exit, for example by mounting thin wires at or near the lip of the jet orifice in~\cite{Gakumase2019Spread} or~\cite{sadeghi2012effects}, or by experimenting with different exit velocity profiles such as a fully developed pipe flow~\cite{xu2002effect}. The calculations also agree with the predictions in~\cite{george1989self}, that describes the dependence of the far field properties on the initial jet conditions.

Entrainment understood as mass flow rate increase per unit axial distance is also easily computed by computing the difference in mass flow rate between two adjacent rings. This turns out to be constant, as expected: 
\begin{equation}
    M_m - M_{m+1} = 2 \pi \rho \left ( r_m u_m - r_{m+1} u_{m+1} \right ) \Delta z = \mathrm{constant},
\end{equation}
noting that a constant inflow of additional mass is required in order to maintain a constant mass density and pressure. The additional mass must come from influx of fluid from the surrounding quiescent fluid pulled in by a lowering of pressure as the jet expands, see e.g.~\cite{ricou1961measurements} and an investigation of the detailed interaction between the jet and the surrounding fluid by DNS~\cite{prabhakaran2019dns}.

\begin{figure*}[t!]
    \centering
    \begin{subfigure}{0.5\textwidth}
        \centering
        \includegraphics[width=\linewidth]{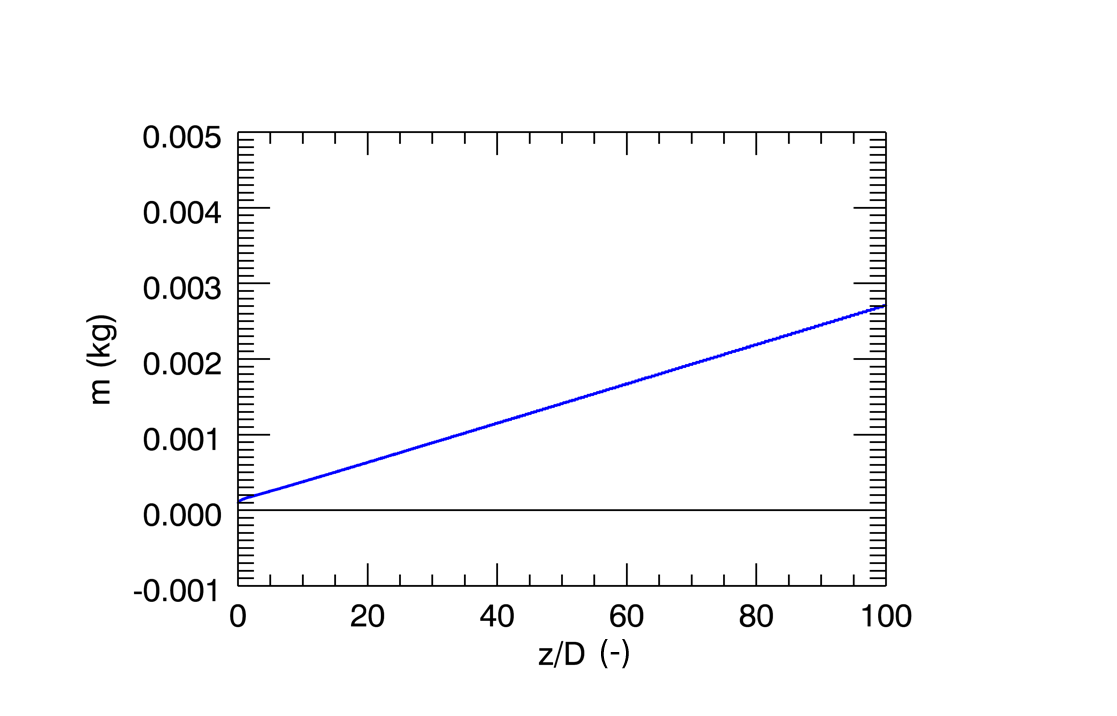}
    \end{subfigure}%
    ~ 
    \begin{subfigure}{0.5\textwidth}
        \centering
        \includegraphics[width=\linewidth]{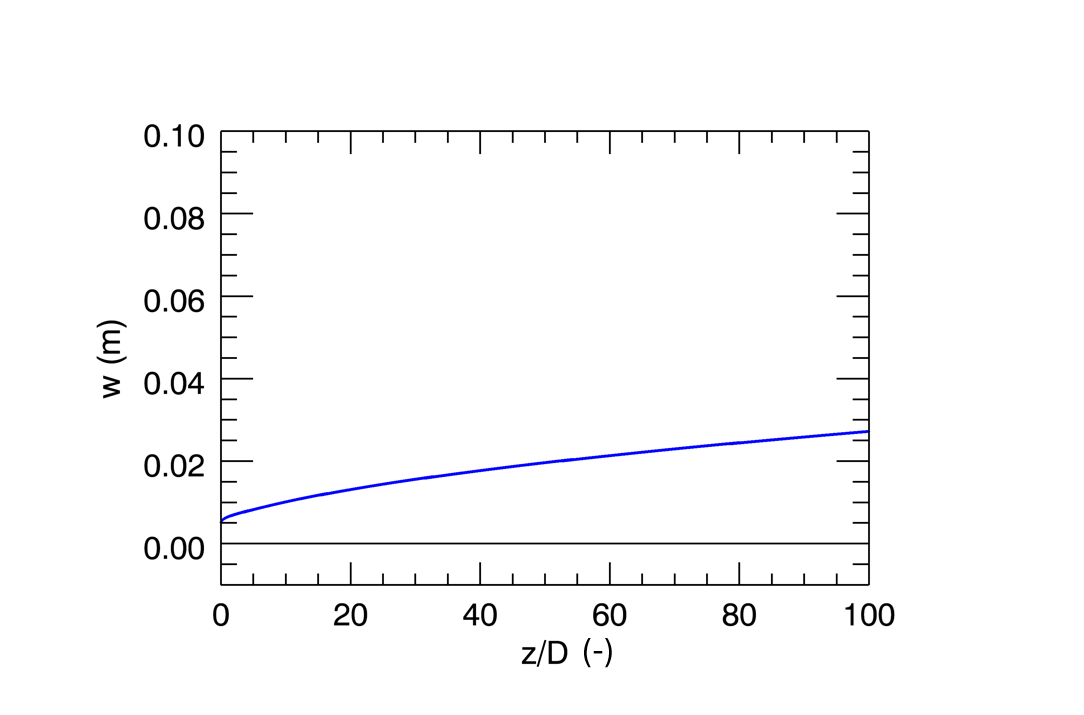}
    \end{subfigure}
    \caption{\label{fig:massandjetwidth} (a) Total mass $m$ of a 10 mm diameter jet with a top-hat initial velocity profile as a function of the jet exit diameter normalized centerline distance $z/D$. (b) Corresponding jet width, measured by the 50\% contour.}
\end{figure*}

In Figure~\ref{fig:massandjetwidth}, we display the linear growth of total mass of the jet within the limit set by 10\% of the centerline velocity. The entrainment is computed with the recursive program for an example of a 10 mm diameter, top-hat jet. Figure~\ref{fig:massandjetwidth}(a) shows the entrainment in the developing region up to 100D. Figure~\ref{fig:massandjetwidth}(b) shows the spreading of the jet measured by the 50\% contour. The interesting thing about these plots is that they display the properties of the jet all the way from the exit to beyond the self-similar region only based on conserved properties resulting from fundamental symmetry properties of space and time.


The conservation of momentum transport rate is easily computed by summing the momentum transport in all the ring elements. Table~\ref{tab:table1} shows the computed momentum transport rate at the jet exit and after $100\,000$ time steps for the example jet used above for different values of the turbulent viscosity.

\begin{table}
    \centering
        \caption{Total momentum transport rate for different turbulent viscosity values.}
    \begin{tabular}{|c|ccc|}\hline
        Turbulent viscosity, $\nu_T$ (m$^2$s$^{-1}$) & 1 & 10 & 100\\\hline
        Initial momentum transport & 0.011569371 & 0.011569371 &	0.011569371 \\\hline
        Final momentum transport & 0.011569365 &	0.011567549	& 0.011565798 \\\hline
    \end{tabular}
    \label{tab:table1}
\end{table}

\subsection{Effect of turbulent viscosity on the far field}

\begin{figure}
\centering
\includegraphics[width=0.5\linewidth]{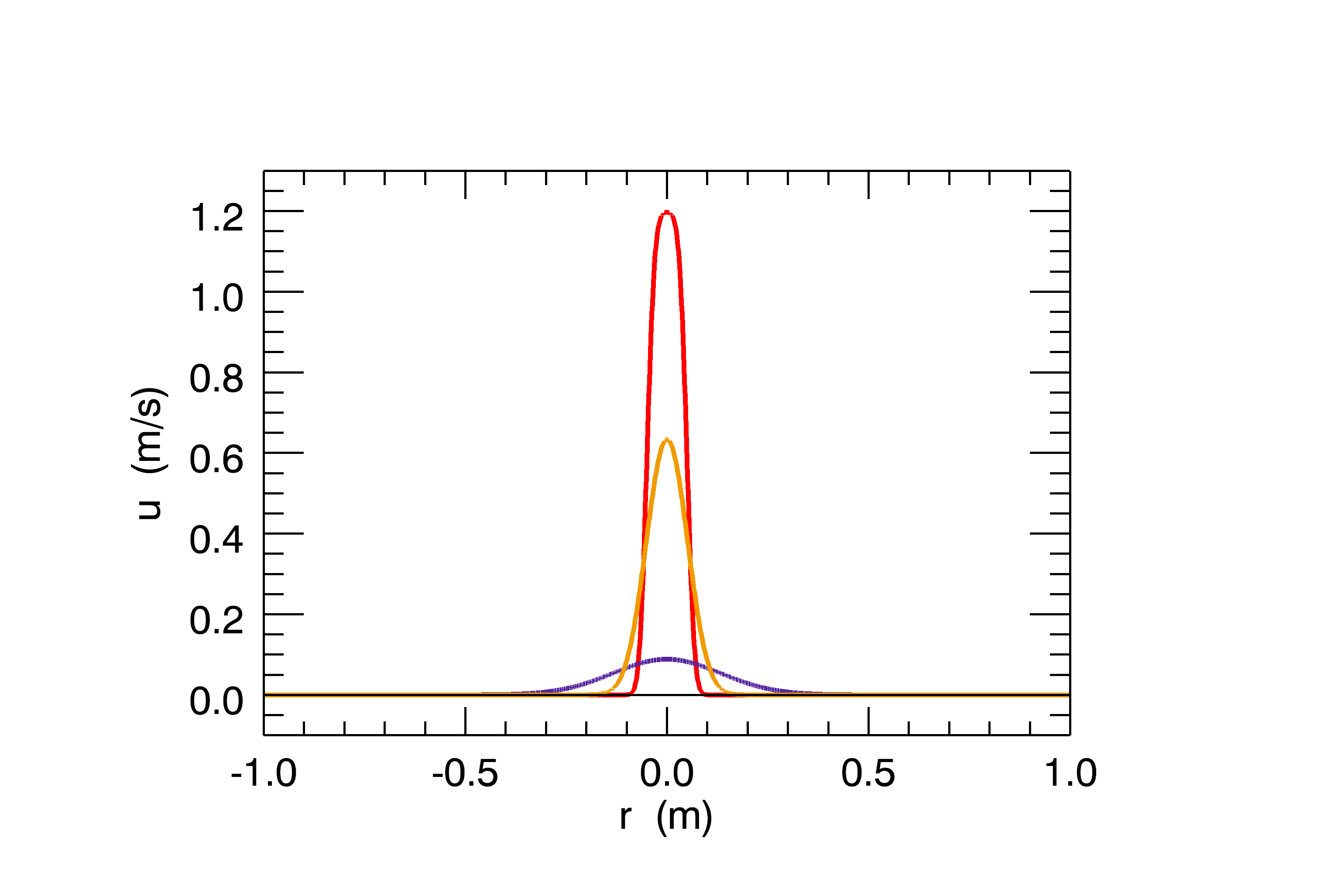}
\caption{\label{fig:turbvisc} Far-field mean velocity at $z/D=100$ for different values of total viscosity. Red: $\nu_{Total} = 1$ m$^2$s$^{-1}$, Yellow: $\nu_{Total} = 10$ m$^2$s$^{-1}$, Blue: $\nu_{Total} = 100$ m$^2$s$^{-1}$.}
\end{figure}

The effect of turbulent viscosity is clearly observed in the plot of the far field velocity profile in Figure~\ref{fig:turbvisc}, where the axial mean velocity is plotted at a distance of 100D for different values of the turbulent viscosity. 

\section{Experiments}\label{sec:experiments}

In order to compare theoretical results to detailed measurements all the way from the jet exit and throughout the self-similar region within the laboratory environment, we used three different size jets with exit diameters $D=\,$100 mm, 50 mm and 10 mm, respectively. All jets were designed with an inner surface profile using an optimized 5$^{th}$ order polynomial to obtain a top-hat initial velocity profile with a minimum surface boundary layer and no separation according to the method described in~\cite{JungThesis}.

\subsection{Hot-wire measurements in the near field of a 100 mm diameter jet}

\begin{figure}
\centering
\includegraphics[width=0.35\linewidth]{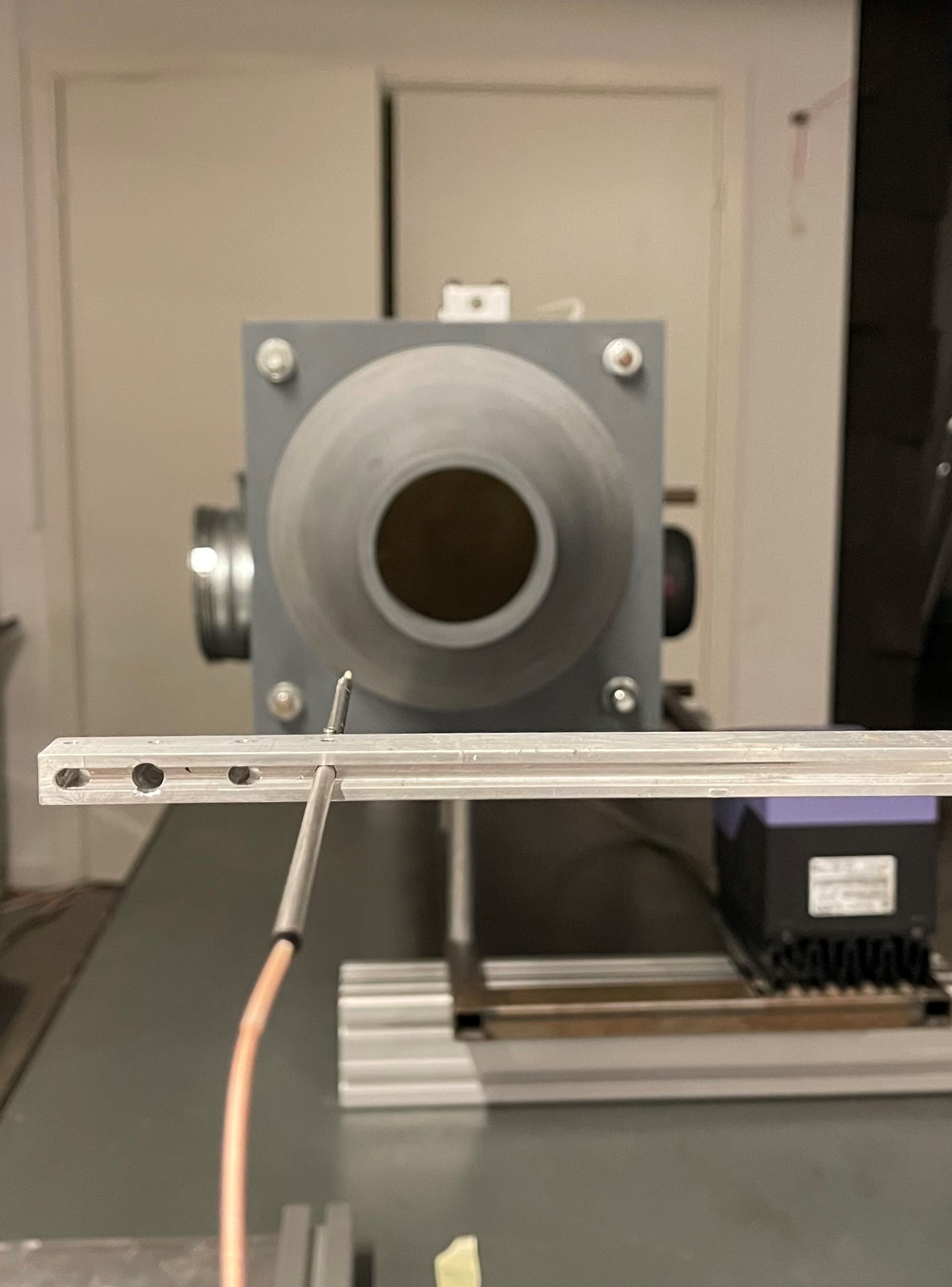}
\caption{\label{fig:100mmjetsetup} Experimental setup for the 100 mm diameter jet measurements.}
\end{figure}

A \SI{5}{\micro\meter} long and 1 mm thick hot-wire (Dantec 55P11) was mounted in the horizontal center plane of the 100 mm diameter jet, as shown in Figure~\ref{fig:100mmjetsetup}. Data was digitized and sampled by a 5444D Picoscope and transferred to a computer for processing.

\begin{figure*}[!h]
    \centering
    \begin{subfigure}{0.5\textwidth}
        \centering
        \includegraphics[width=\linewidth]{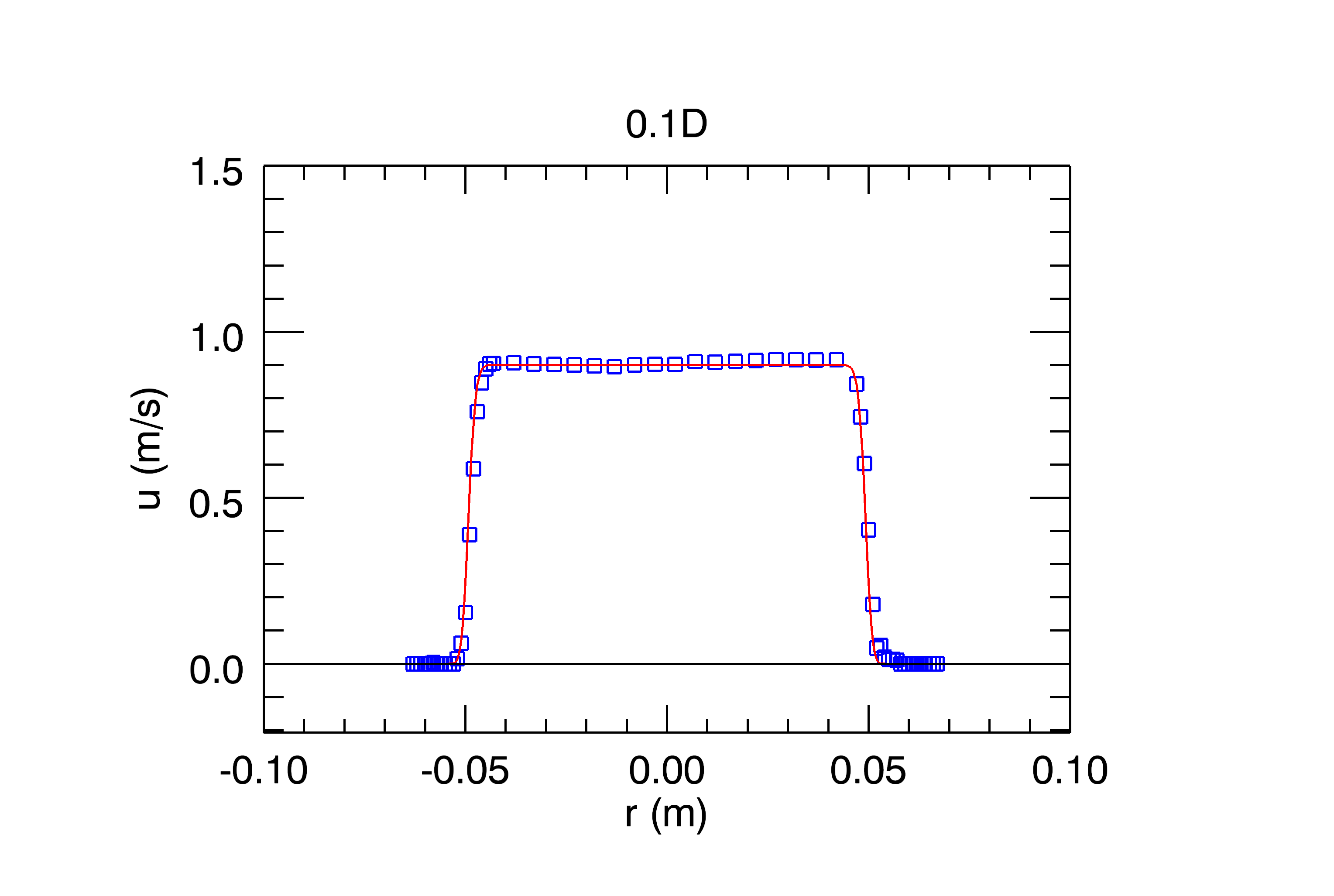}
    \end{subfigure}%
    ~ 
    \begin{subfigure}{0.5\textwidth}
        \centering
        \includegraphics[width=\linewidth]{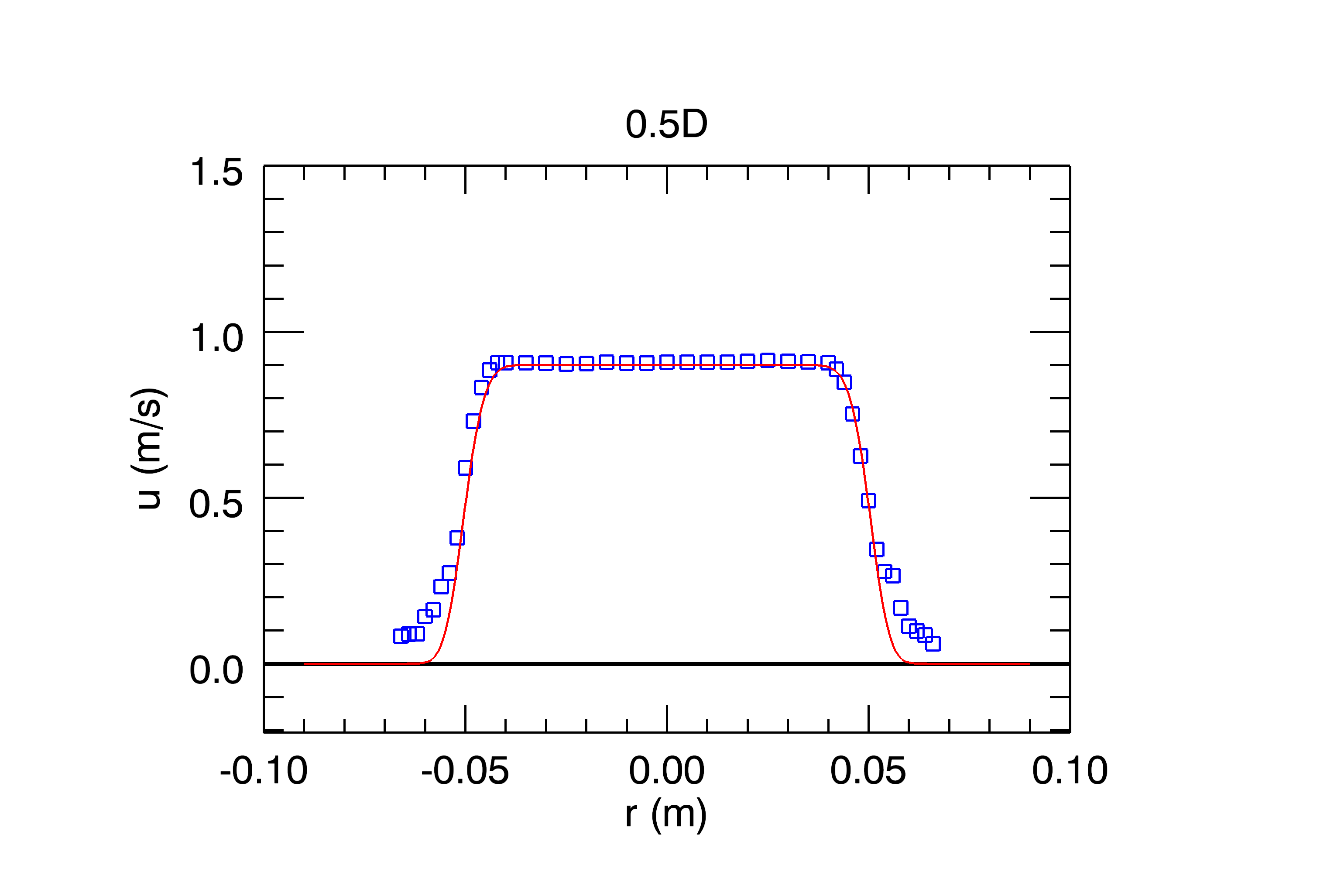}
    \end{subfigure}
    \begin{subfigure}{0.5\textwidth}
        \centering
        \includegraphics[width=\linewidth]{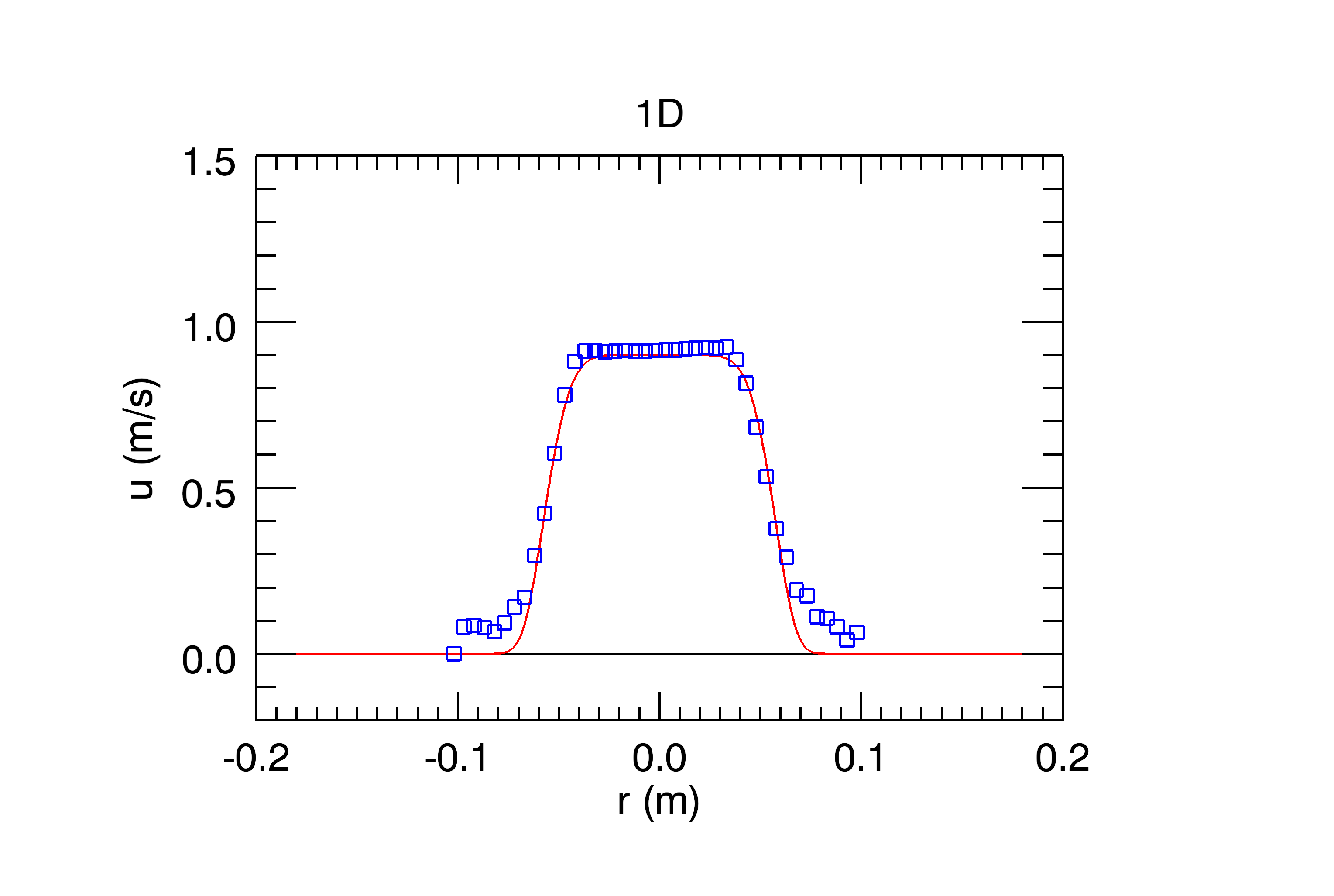}
    \end{subfigure}
    ~ 
    \begin{subfigure}{0.46\textwidth}
        \centering
        \includegraphics[width=\linewidth]{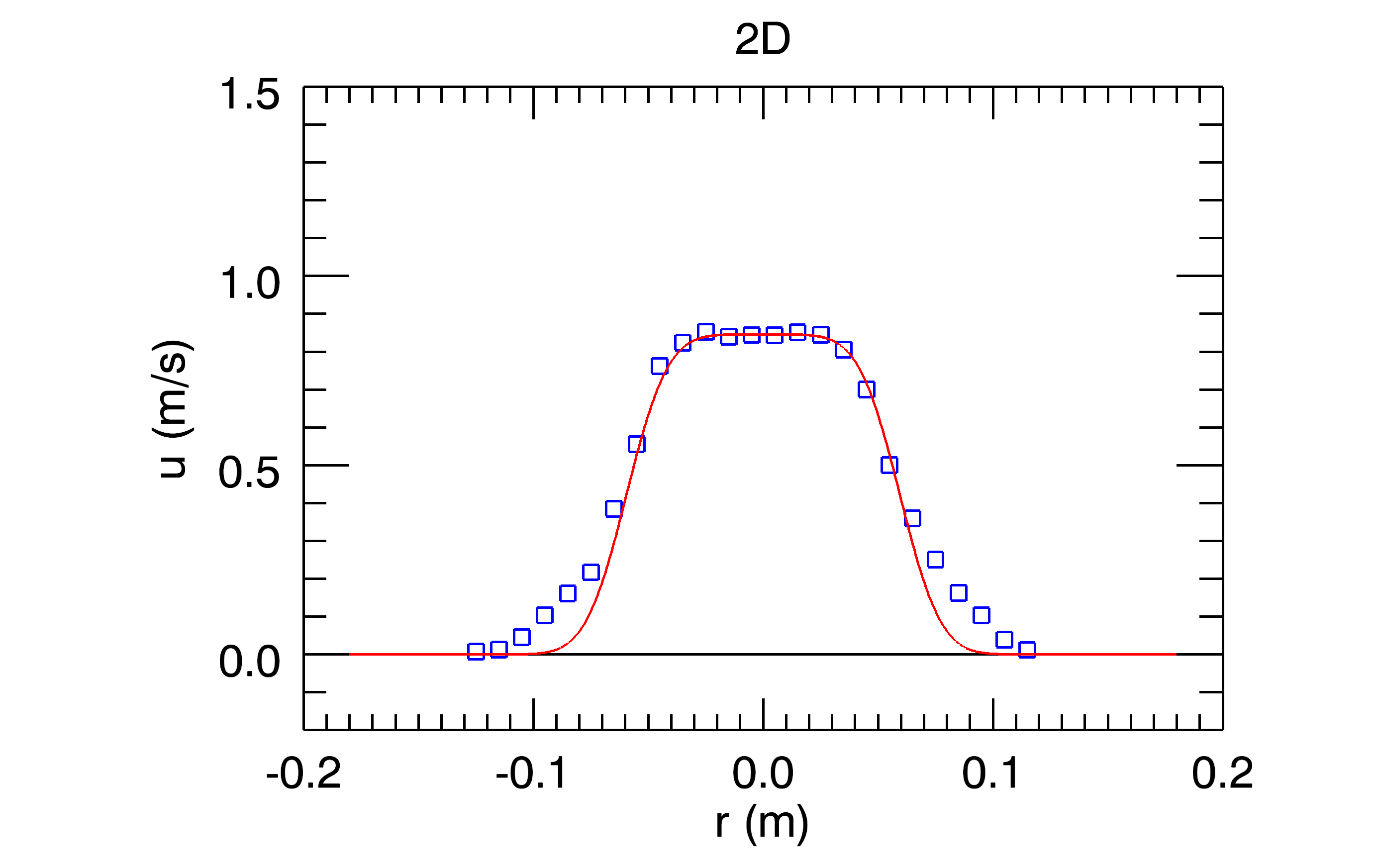}
    \end{subfigure}
    \caption{\label{fig:meas100mmjet} Measured averaged velocities $u$ from a single hotwire (blue squares) at $z/D=\,$ 0.1, 0.5, 1 and 2 as a function of radial position, $r$, overlayed with numerically computed profiles (red curves).}
\end{figure*}

Figures~\ref{fig:meas100mmjet}a-d show measured mean velocity profiles at 0.1D, 0.5D, 1D and 2D overlayed with computed profiles with 1, 40, 150 and 275 time steps. Evidently, the computed profiles match well with the measured mean velocity profiles in the developing part of the jet. The discrepancy in the outer part is due to the inability of a hot-wire to measure high intensity turbulence correctly.

\subsection{Laser Doppler anemometry (LDA) measurements in a 50 mm jet}

\begin{figure}[!h]
\centering
\includegraphics[width=0.45\linewidth]{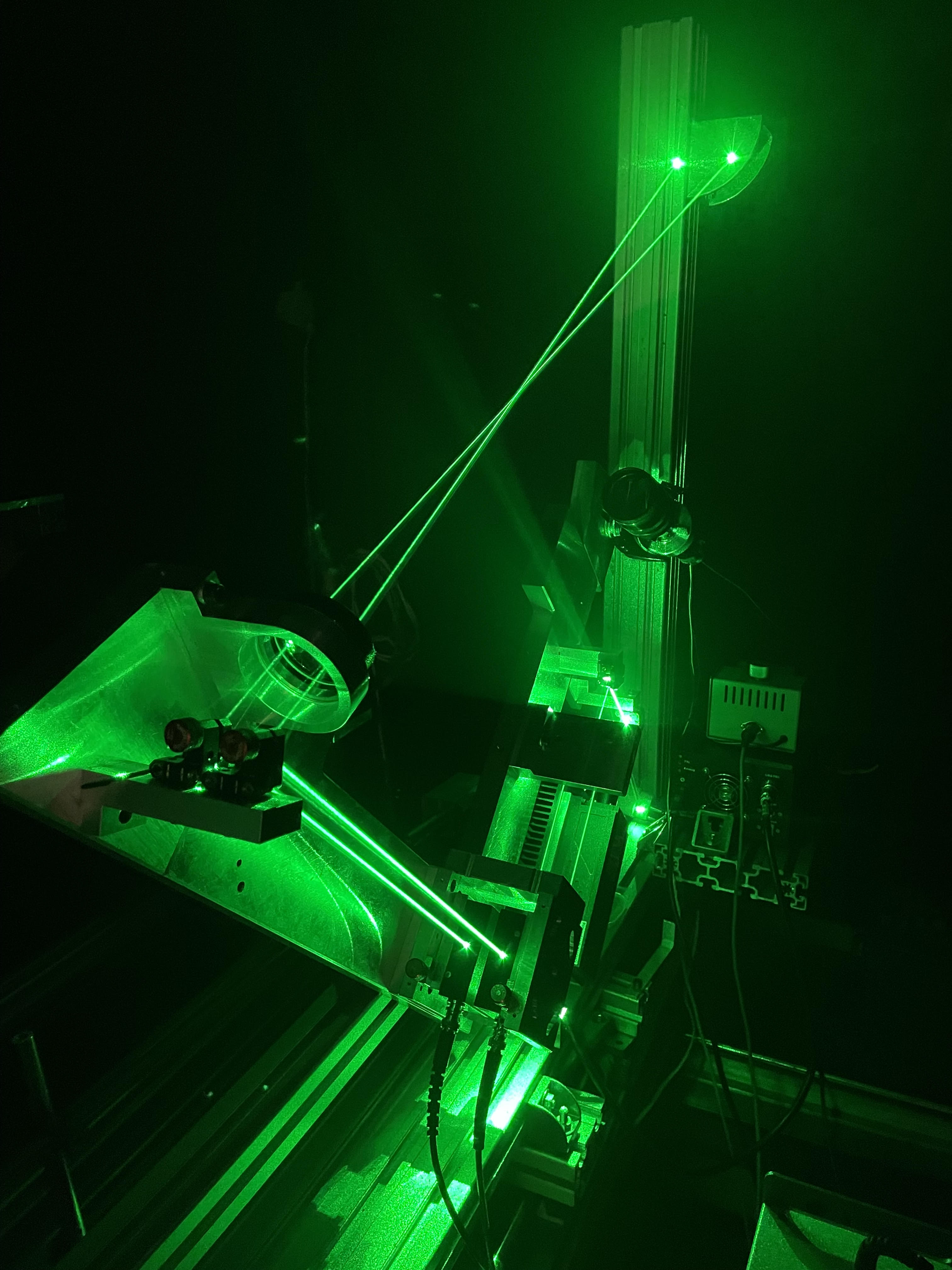}
\caption{\label{fig:meas50mmjet} LDA optics.}
\end{figure}

Measurements in the near field of a 50 mm diameter jet were performed with an in-house designed software-driven LDA~\cite{yaacob2019novel}. The LDA optics is shown in Figure~\ref{fig:meas50mmjet}. The measurement volume is nearly spherical with a diameter of \SI{300}{\micro\meter} due to the side-scattering detection optics configuration. The jet flow and the volume surrounding the jet were seeded with 1-\SI{4}{\micro\meter} diameter sized glycerin droplets.

The photodetector signal is digitized with a 5444D Picoscope and the signal stored on hard disc in the form of a selected number of records. The records are later processed to provide data in the form of the axial velocity component, particle arrival time and residence time. A dual-Brag cell module allows to introduce an optical frequency shift for optimum, bias-free signal processing. The signal processing is entirely software driven allowing maximum flexibility for extraction of data from the signal and for further processing to obtain low noise, bias free flow statistics.

\begin{figure*}[t!]
    \centering
    \begin{subfigure}{0.49\textwidth}
        \centering
        \includegraphics[width=\linewidth]{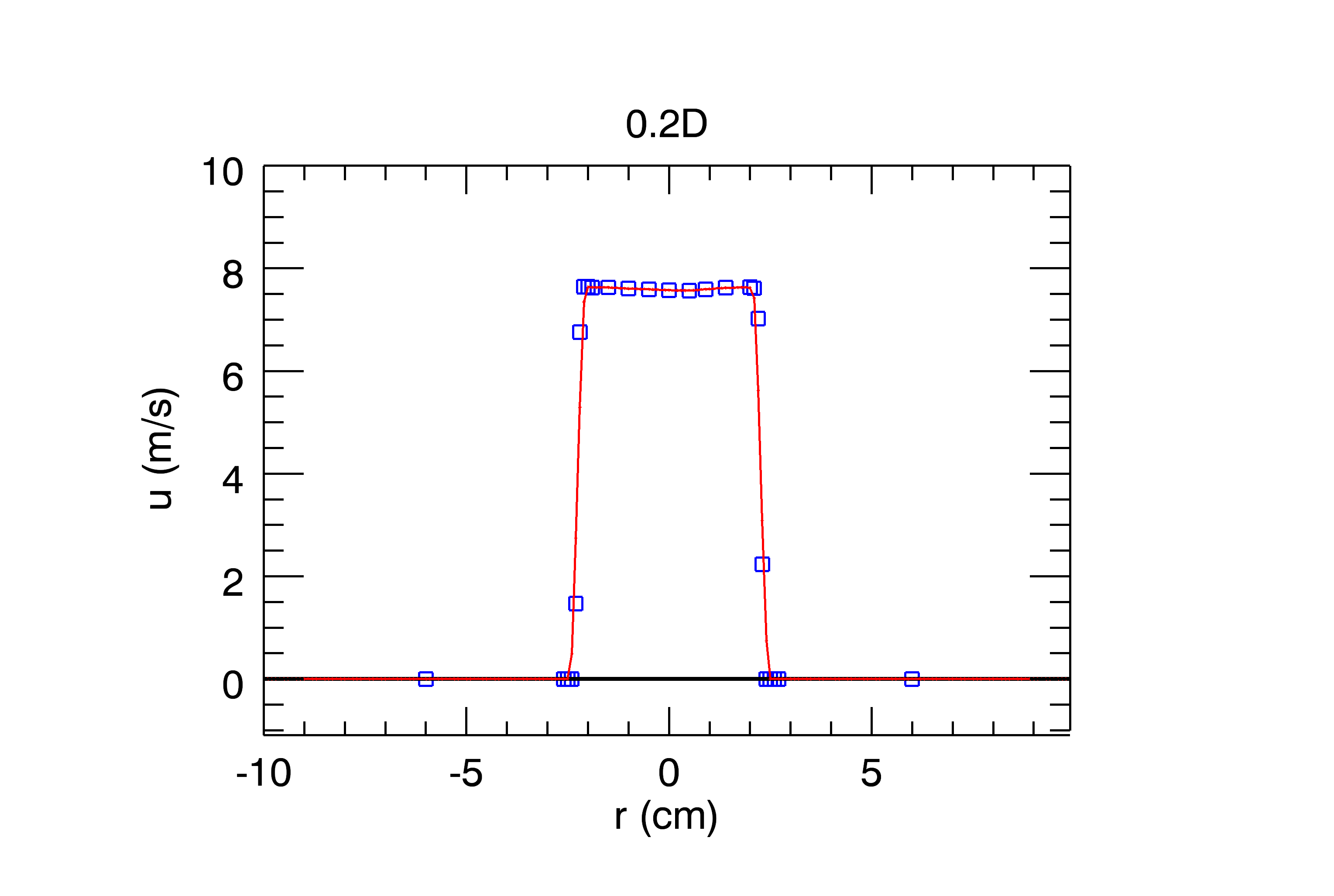}
    \end{subfigure}%
    ~ 
    \begin{subfigure}{0.49\textwidth}
        \centering
        \includegraphics[width=\linewidth]{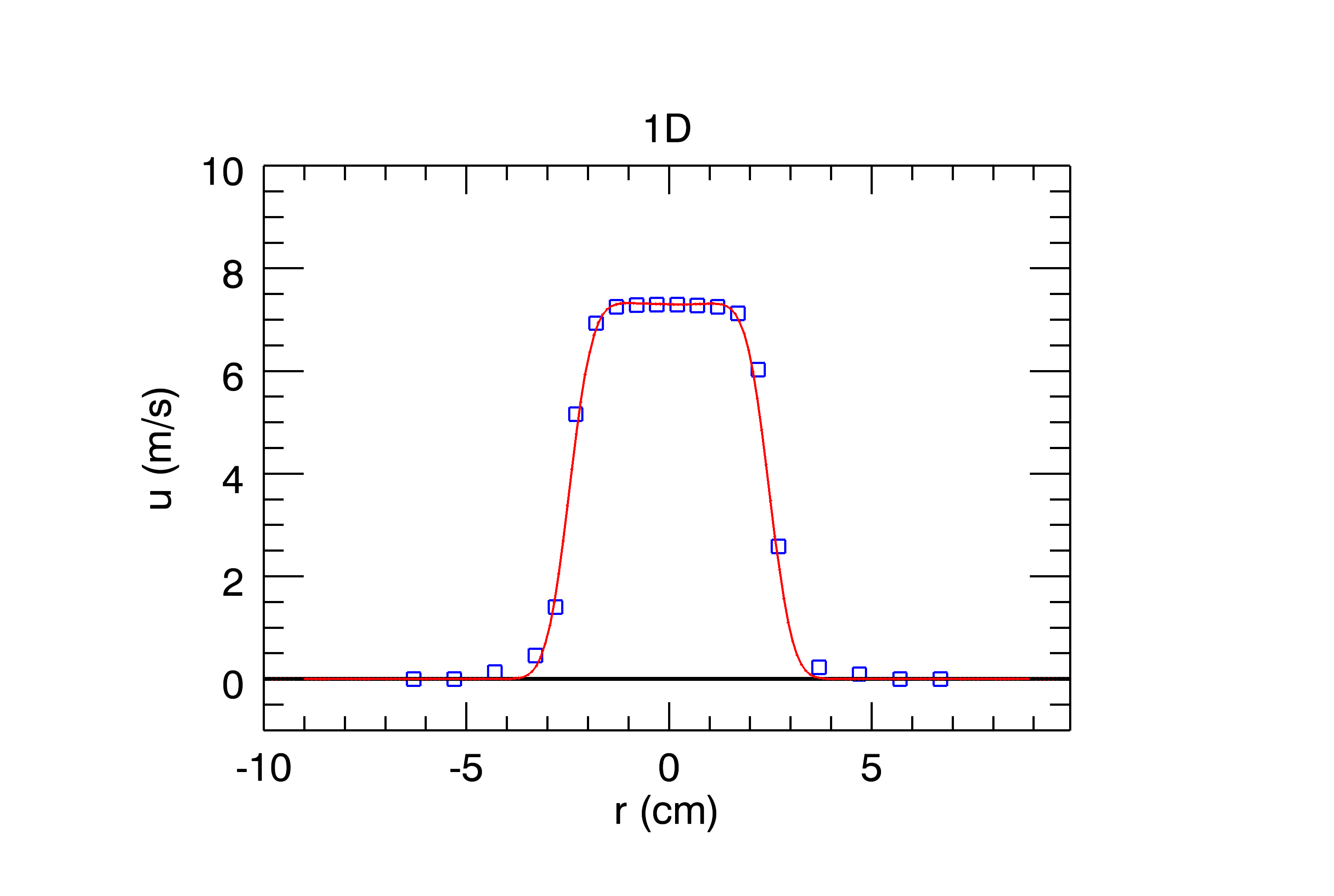}
    \end{subfigure}
    \begin{subfigure}{0.49\textwidth}
        \centering
        \includegraphics[width=\linewidth]{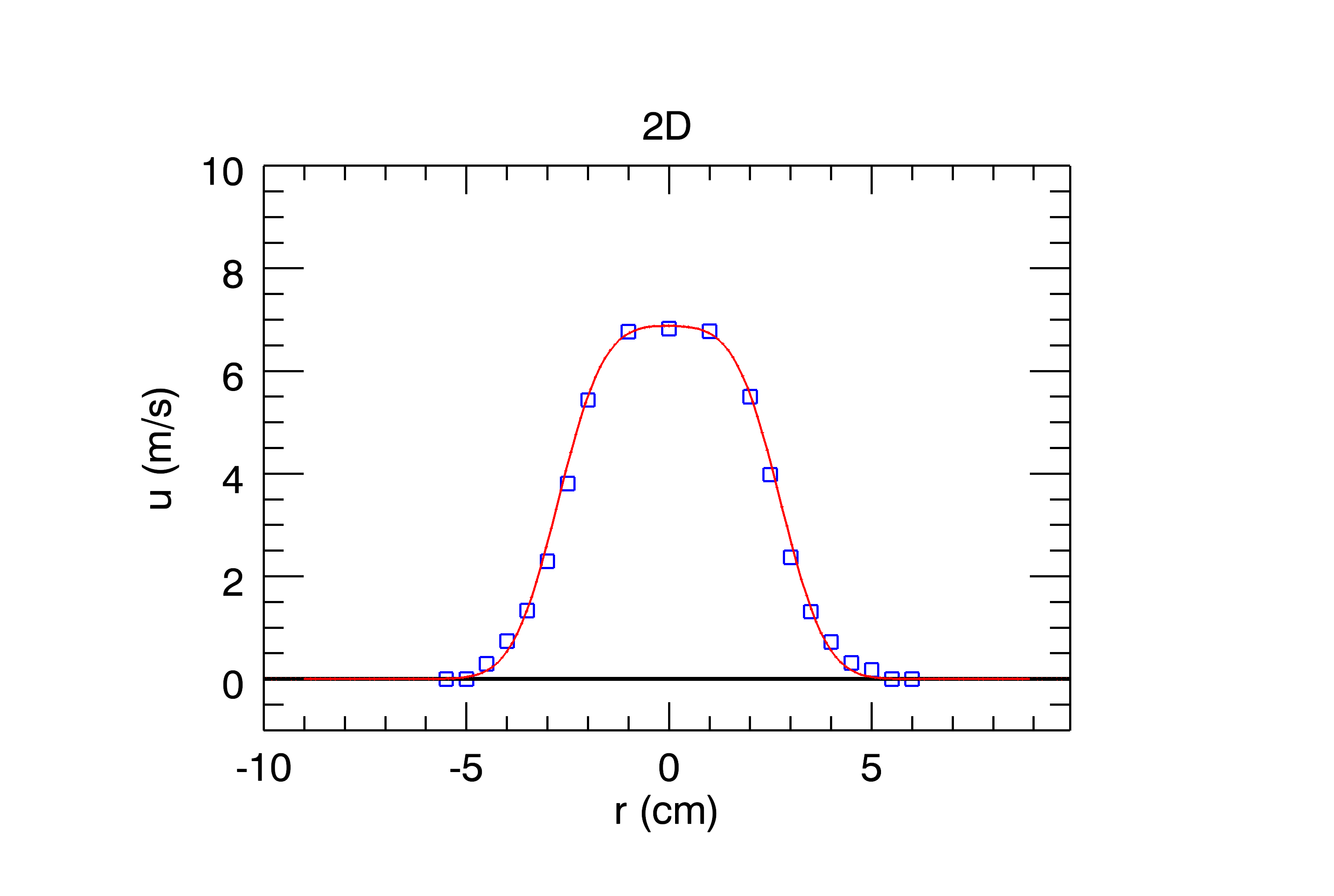}
    \end{subfigure}
    ~ 
    \begin{subfigure}{0.49\textwidth}
        \centering
        \includegraphics[width=\linewidth]{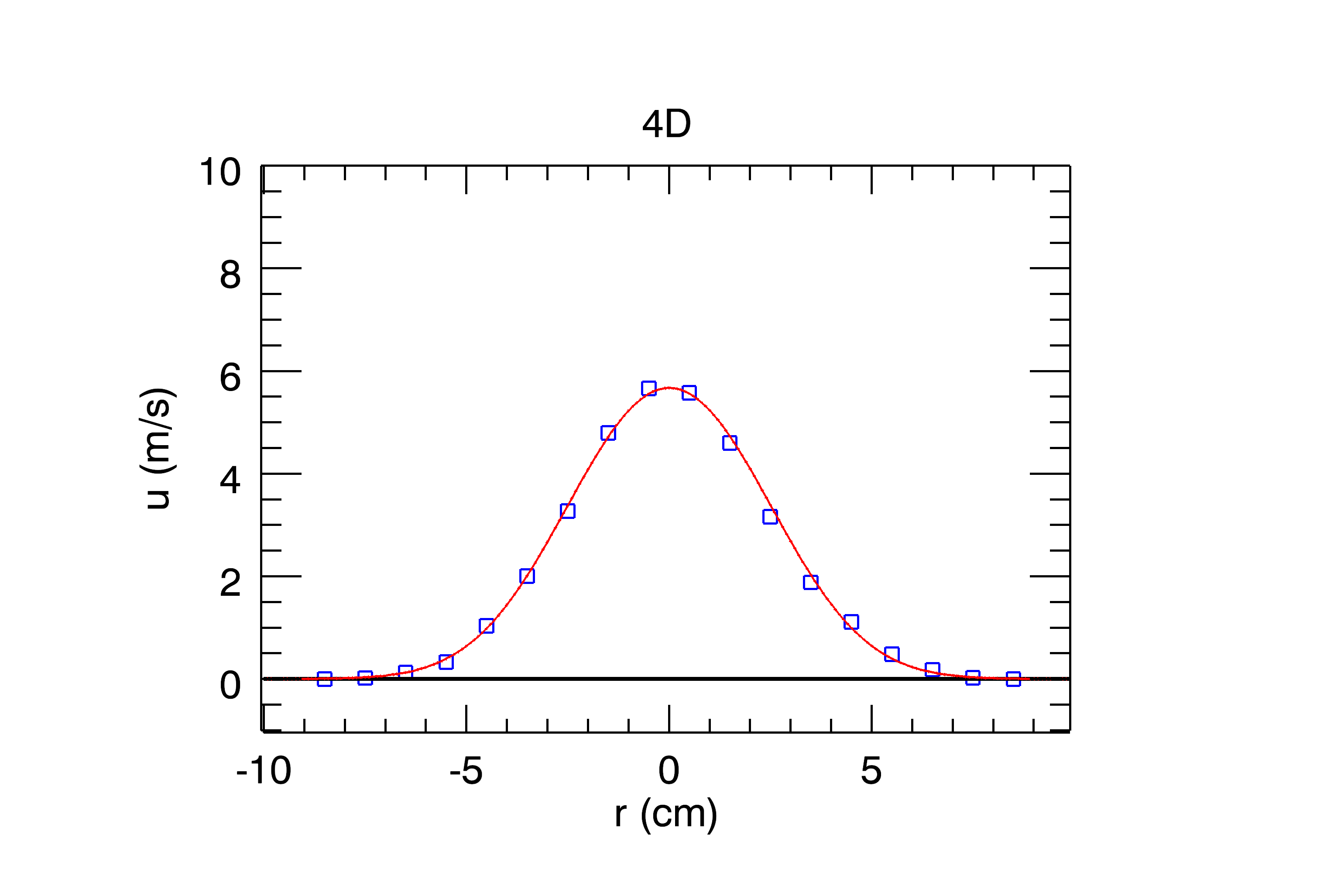}
    \end{subfigure}
        \begin{subfigure}{0.49\textwidth}
        \centering
        \includegraphics[width=\linewidth]{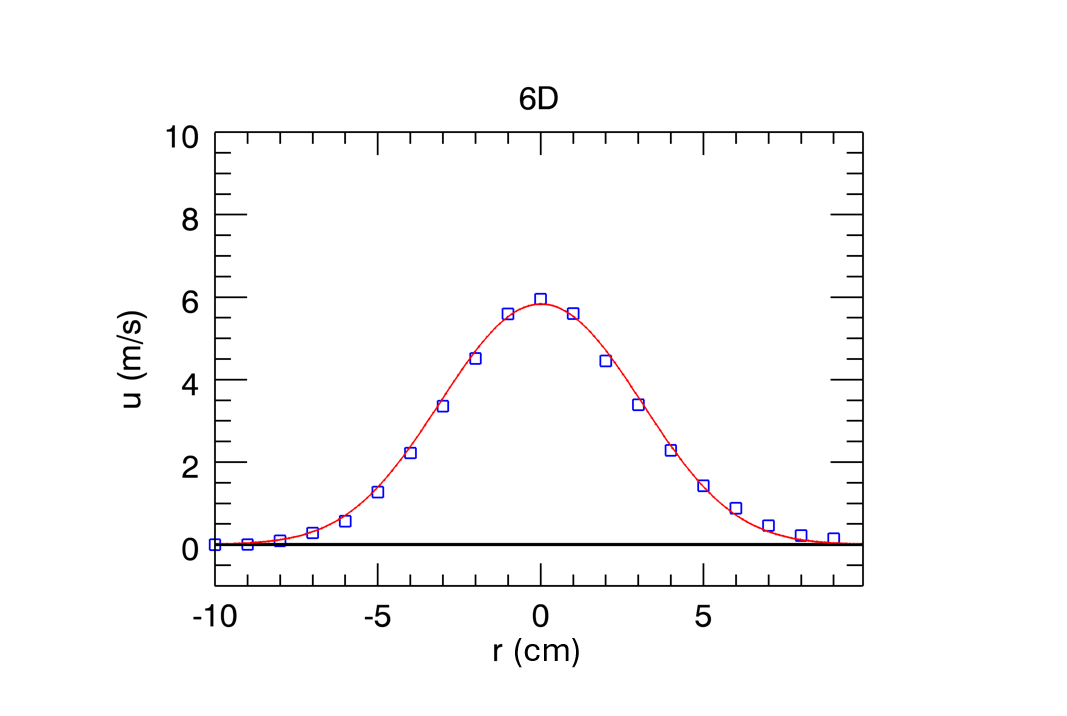}
    \end{subfigure}
    \begin{subfigure}{0.49\textwidth}
        \centering
        \includegraphics[width=\linewidth]{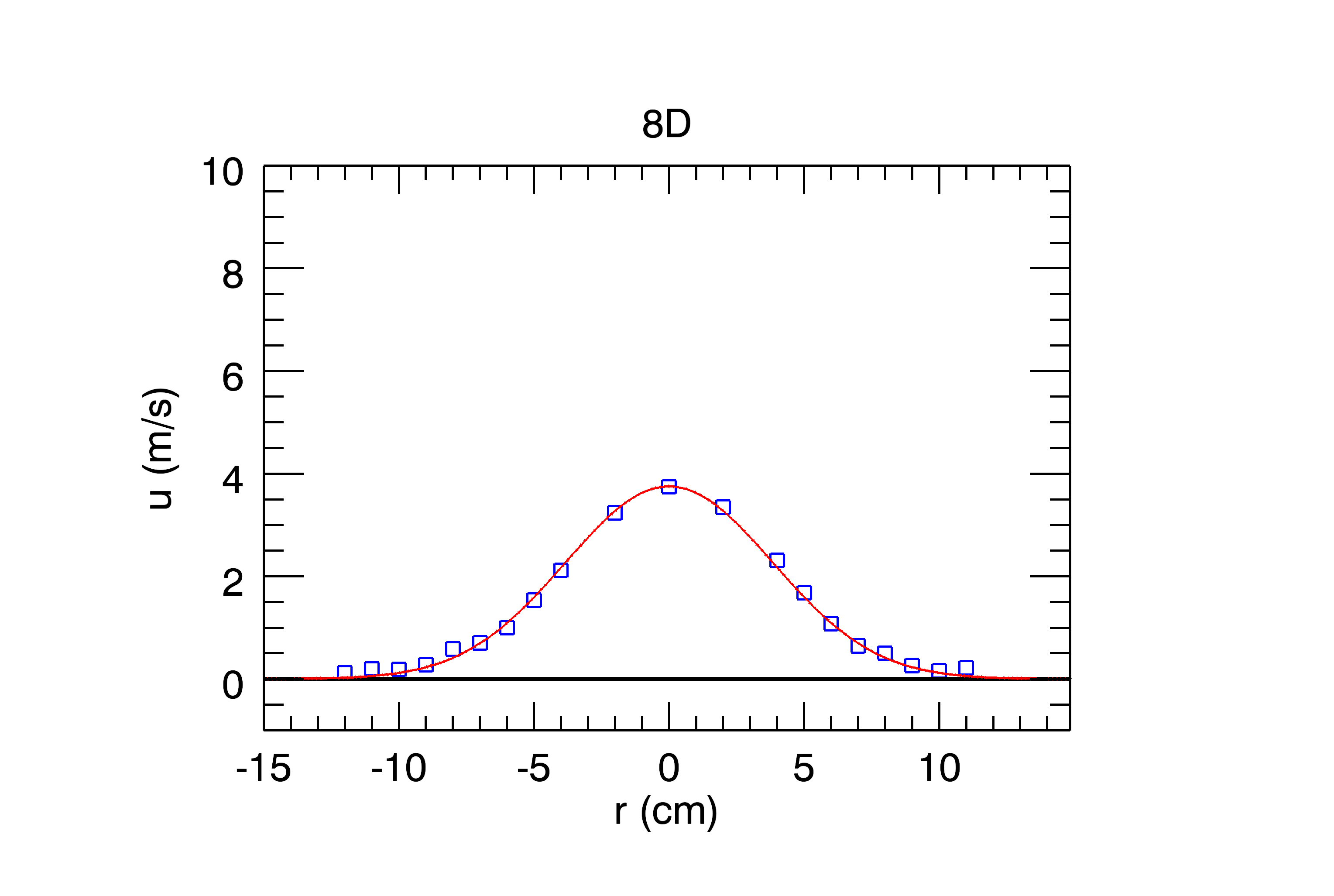}
    \end{subfigure}
    \caption{\label{fig:50mmjetprofiles} Measured near field jet axial velocity profiles (blue squares) as a function of radial position and computed velocity cross section profiles (red curves) in the 50 mm jet. The computed red curve overlays perfectly with the analytical expression (green curve), which is thus difficult to discern.}
\end{figure*}

The cross-section measurements are performed in the near to medium velocity field (0.2D – 8D) of the 50 mm diameter jet at a Reynolds number $Re = 30\,000$ based on the jet exit velocity and diameter. The results are depicted in Figure~\ref{fig:50mmjetprofiles}. The measured velocity profiles (blue squares) are plotted together with data obtained from the recursive algorithm (red curves) and from the analytical expression (green curve). The red and green curves are nearly identical and appear as one in the Figure. The three sets of data show near perfect agreement confirming the correctness of the model based on fundamental symmetries and on Newtonian friction forces.

\newpage


\begin{figure}[!h]
\centering
\includegraphics[width=0.3\linewidth]{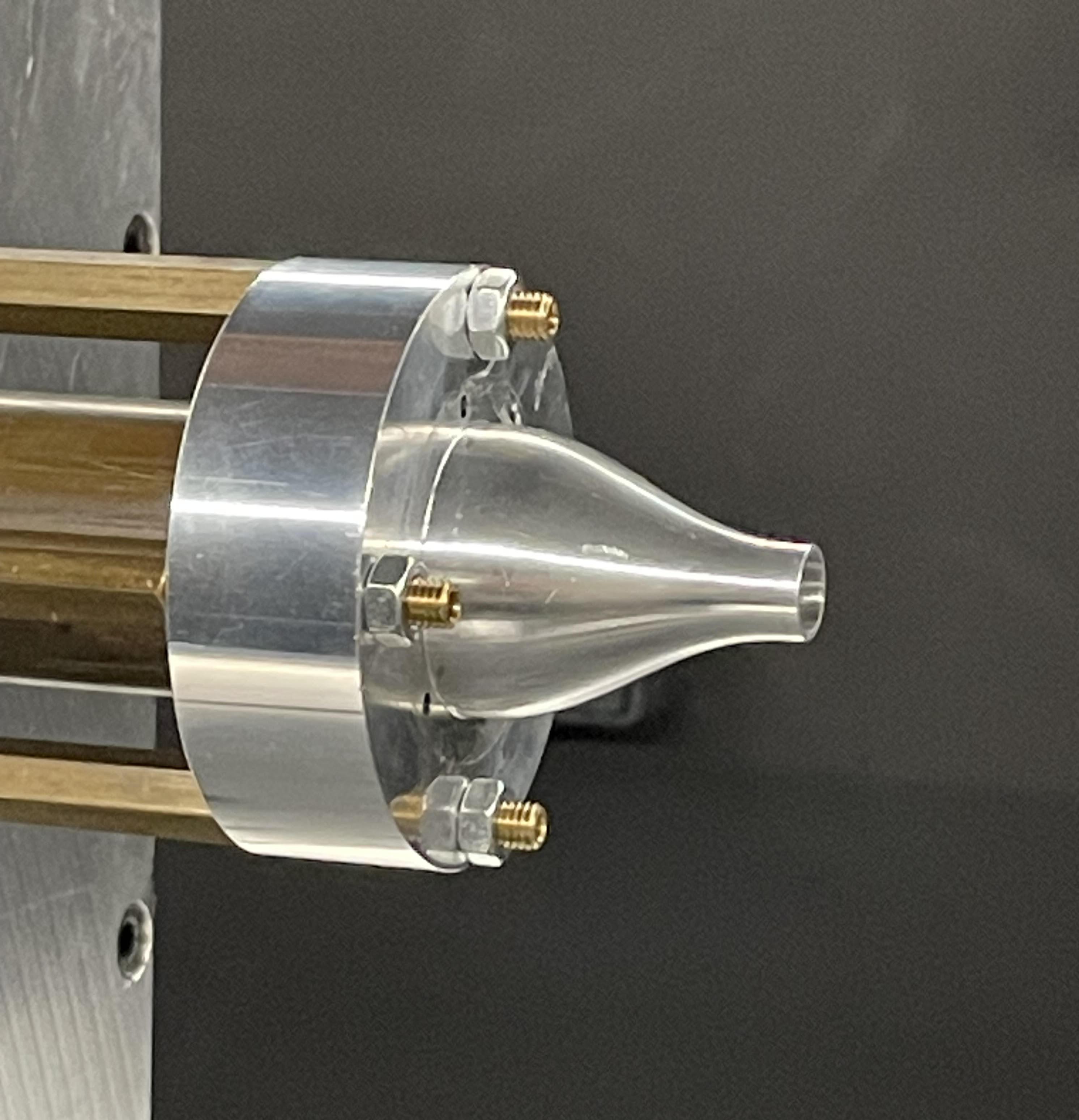}
\caption{\label{fig:10mmjet2} 10 mm jet nozzle.}
\end{figure}

\subsection{Measurements in a 10 mm jet}

The 10 mm diameter jet employed is shown in Figure~\ref{fig:10mmjet2}. Cross section measurements were performed with the LDA at $z/D=\,$30 – 100 with 10$D$ increments.

\begin{figure*}[!h]
    \centering
    \begin{subfigure}{0.49\textwidth}
        \centering
        \includegraphics[width=\linewidth]{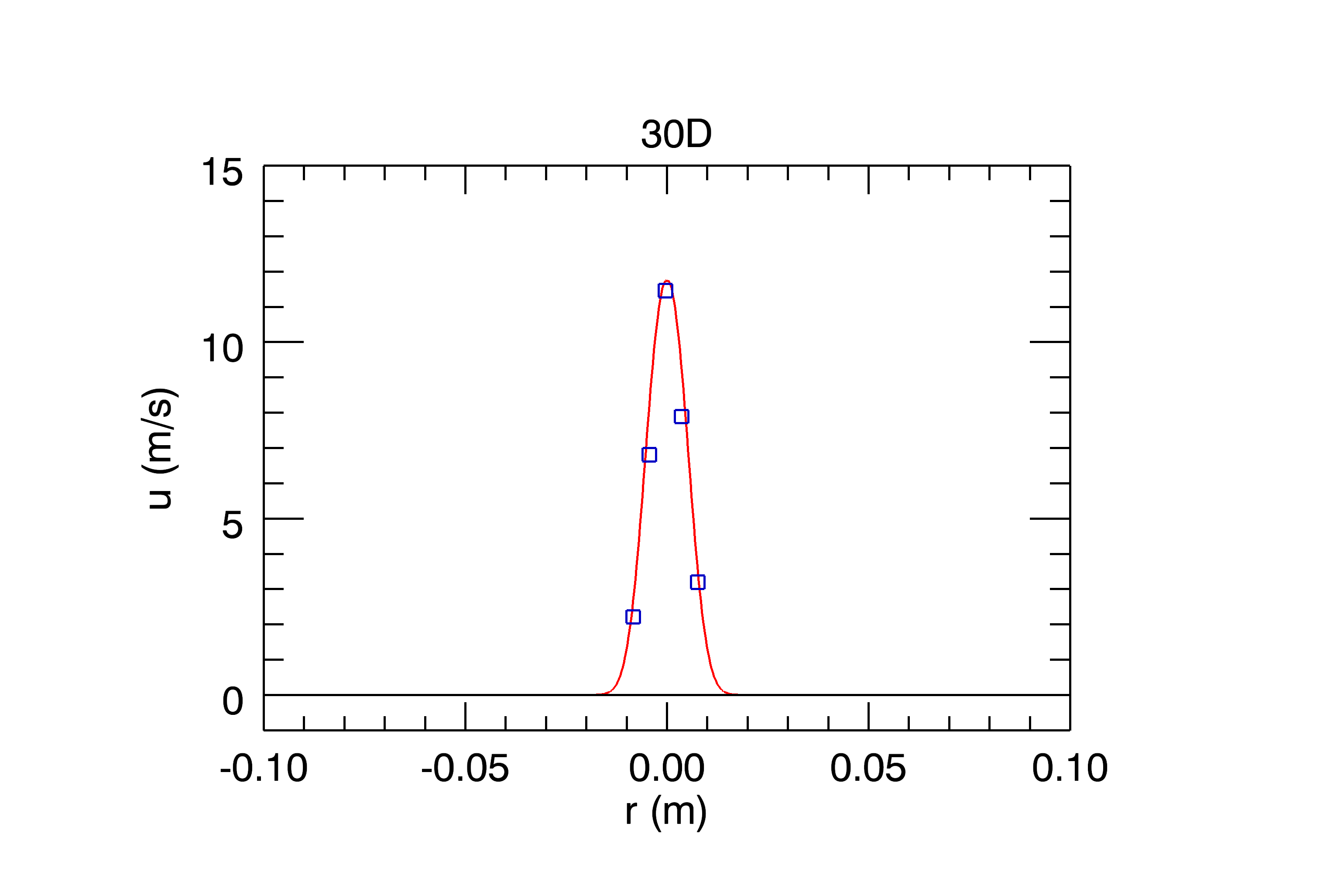}
    \end{subfigure}%
    ~ 
    \begin{subfigure}{0.49\textwidth}
        \centering
        \includegraphics[width=\linewidth]{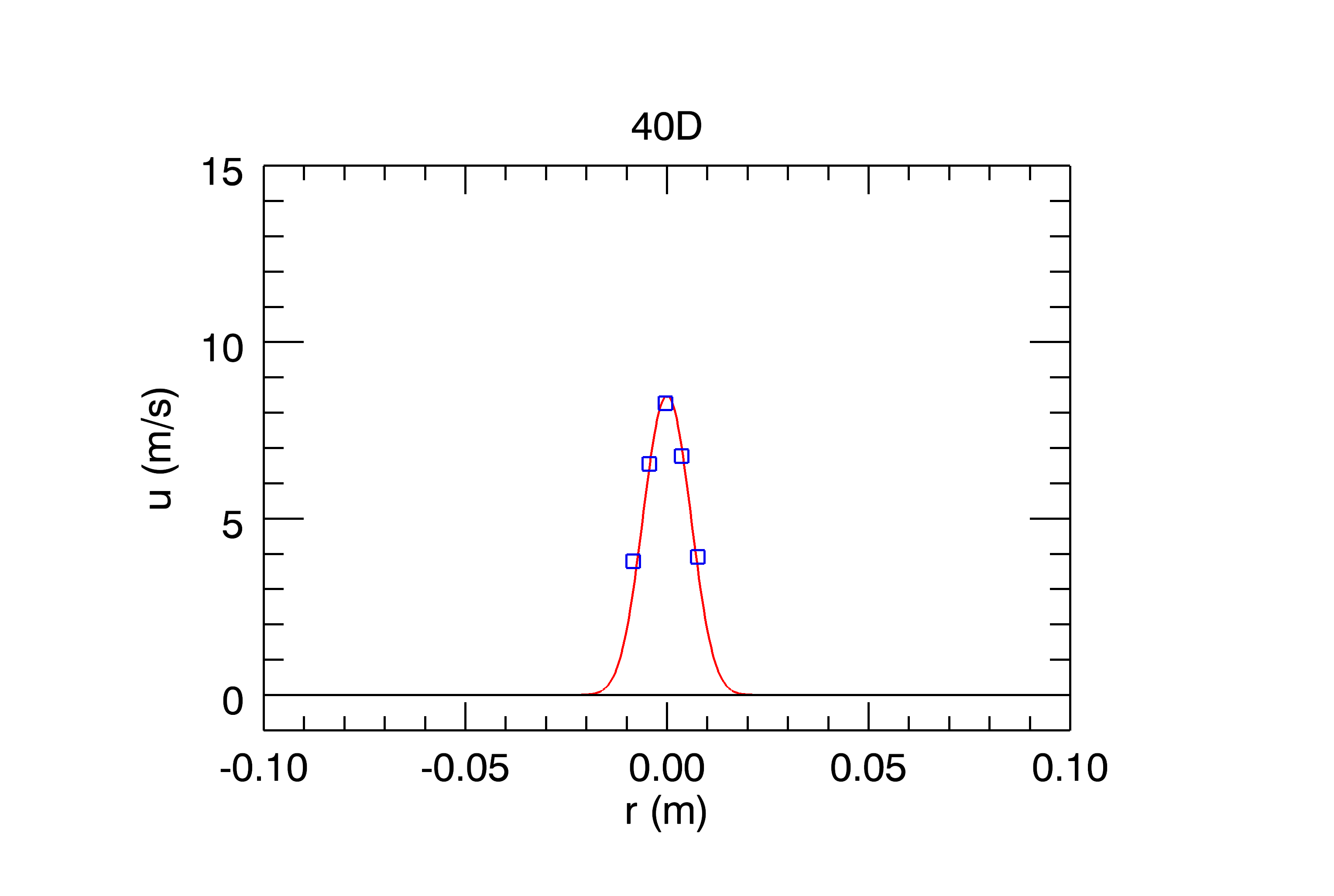}
    \end{subfigure}
    \begin{subfigure}{0.49\textwidth}
        \centering
        \includegraphics[width=\linewidth]{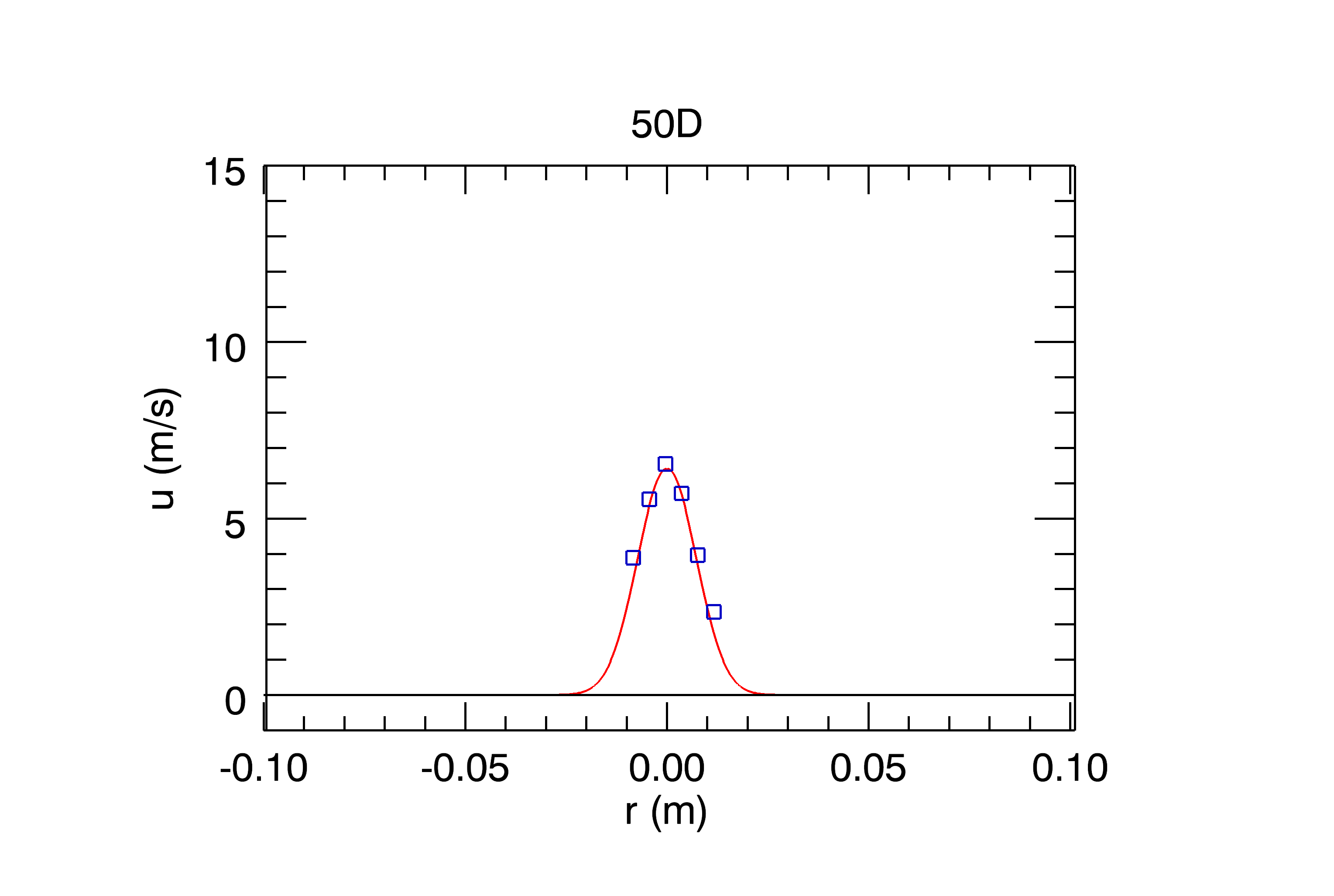}
    \end{subfigure}
    ~ 
    \begin{subfigure}{0.49\textwidth}
        \centering
        \includegraphics[width=\linewidth]{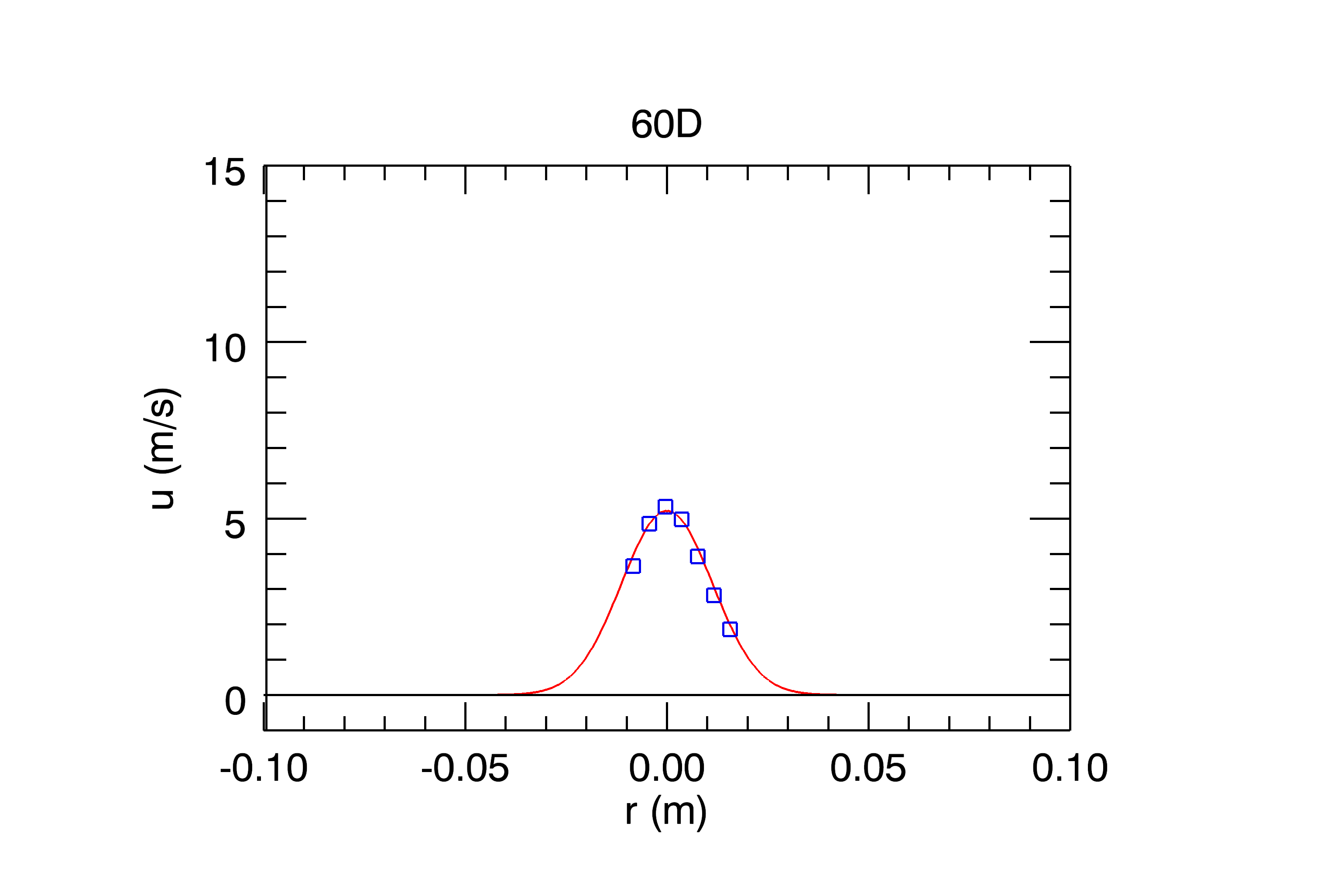}
    \end{subfigure}
        \begin{subfigure}{0.49\textwidth}
        \centering
        \includegraphics[width=\linewidth]{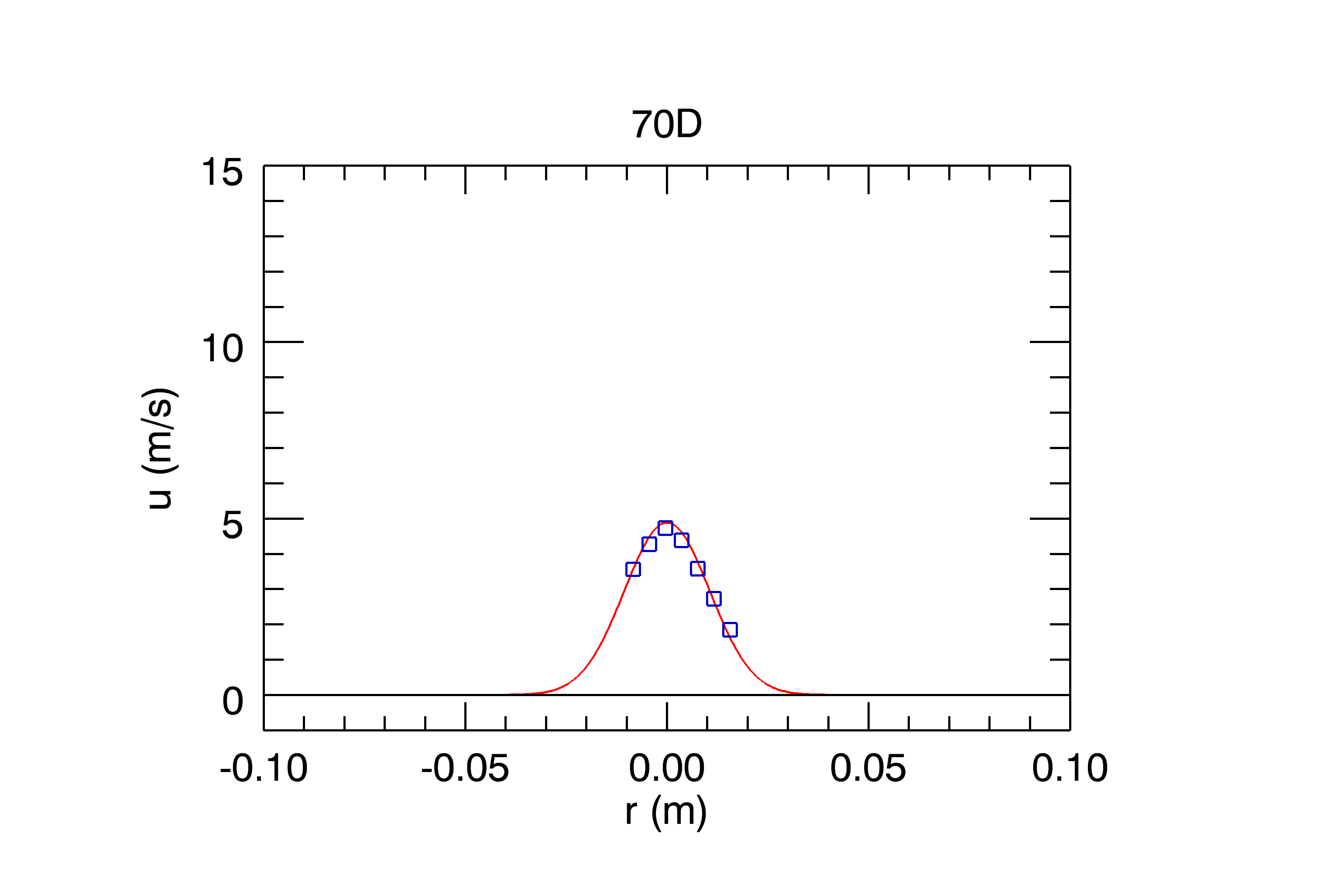}
    \end{subfigure}
    \begin{subfigure}{0.49\textwidth}
        \centering
        \includegraphics[width=\linewidth]{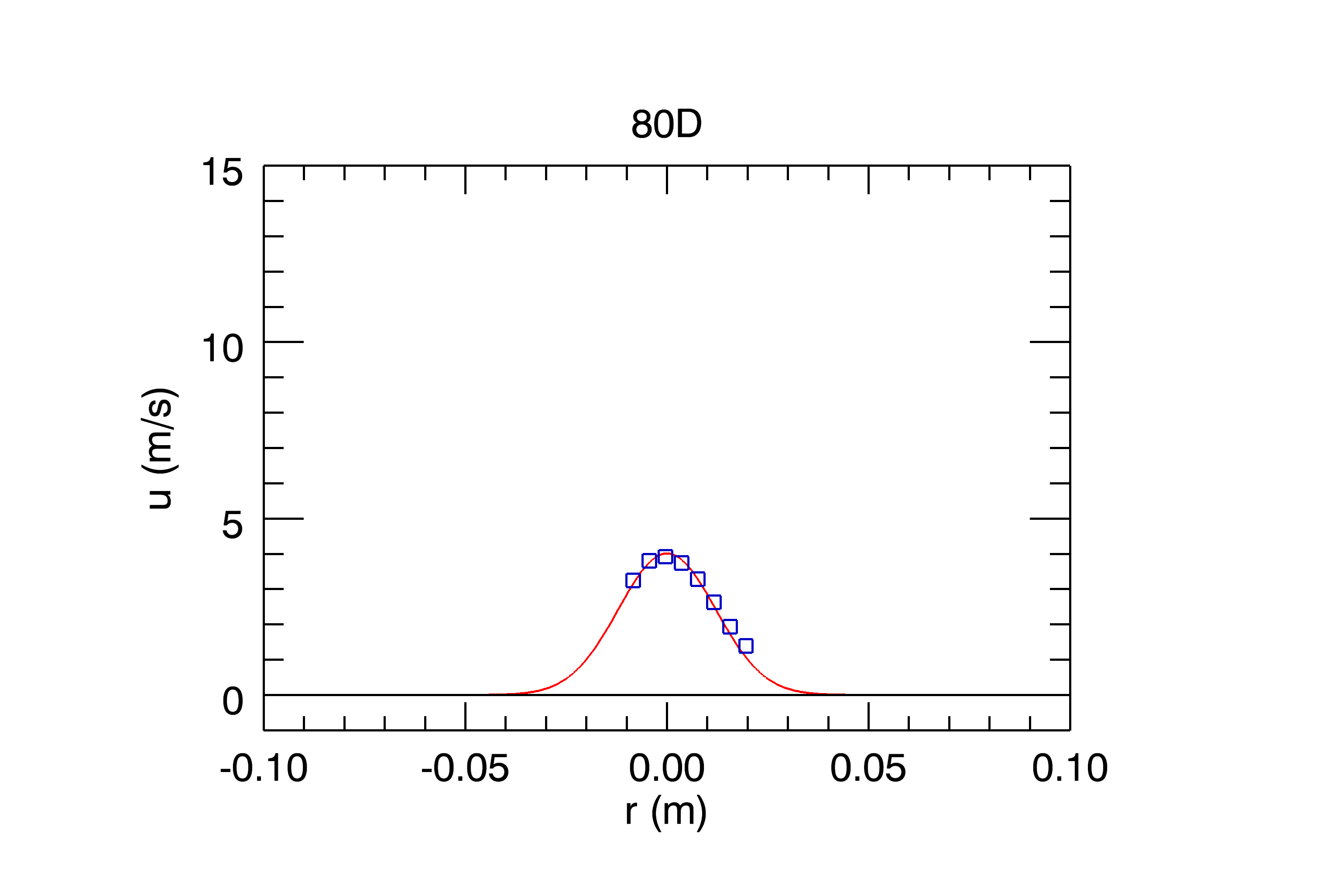}
    \end{subfigure}
            \begin{subfigure}{0.49\textwidth}
        \centering
        \includegraphics[width=\linewidth]{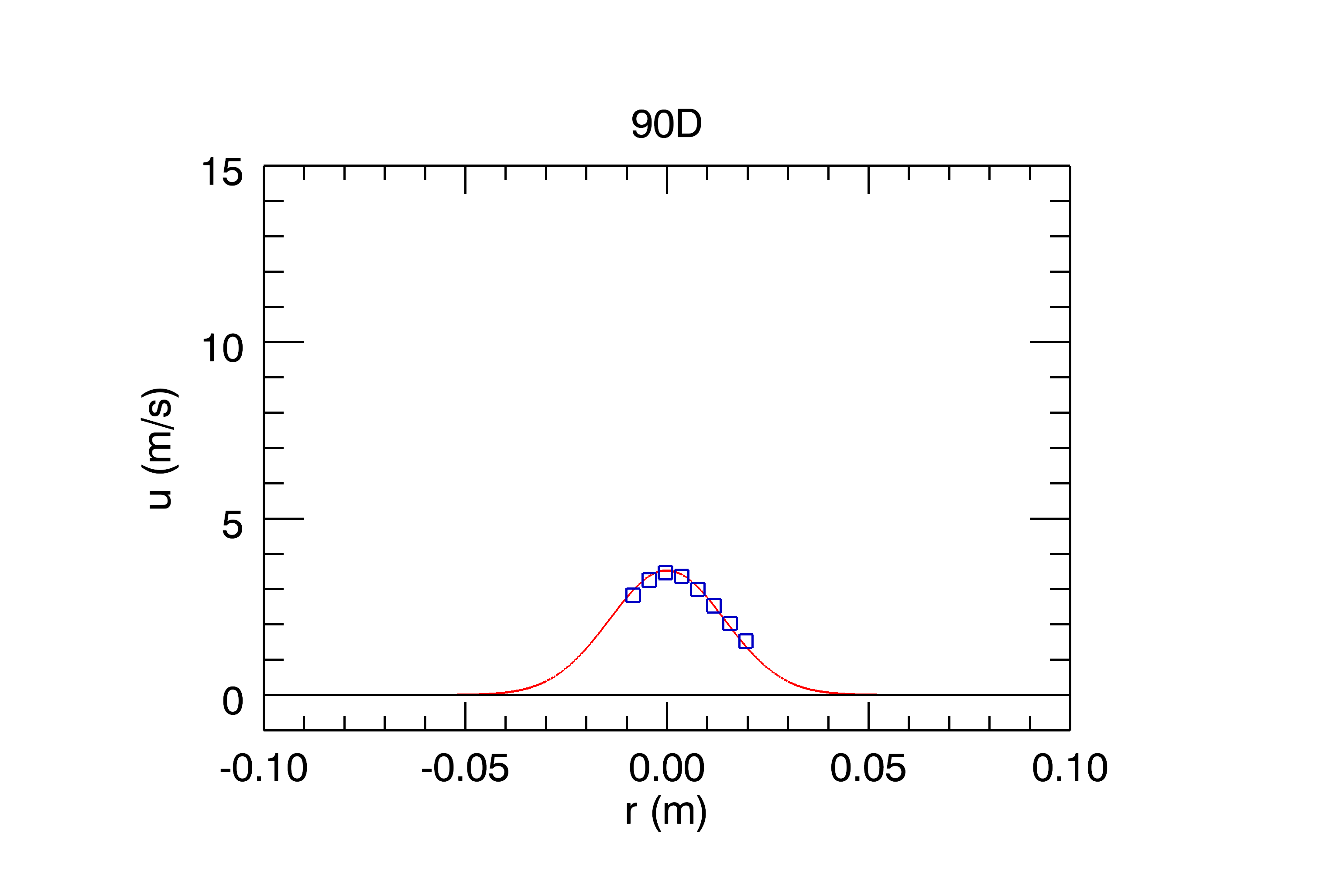}
    \end{subfigure}
    \begin{subfigure}{0.49\textwidth}
        \centering
        \includegraphics[width=\linewidth]{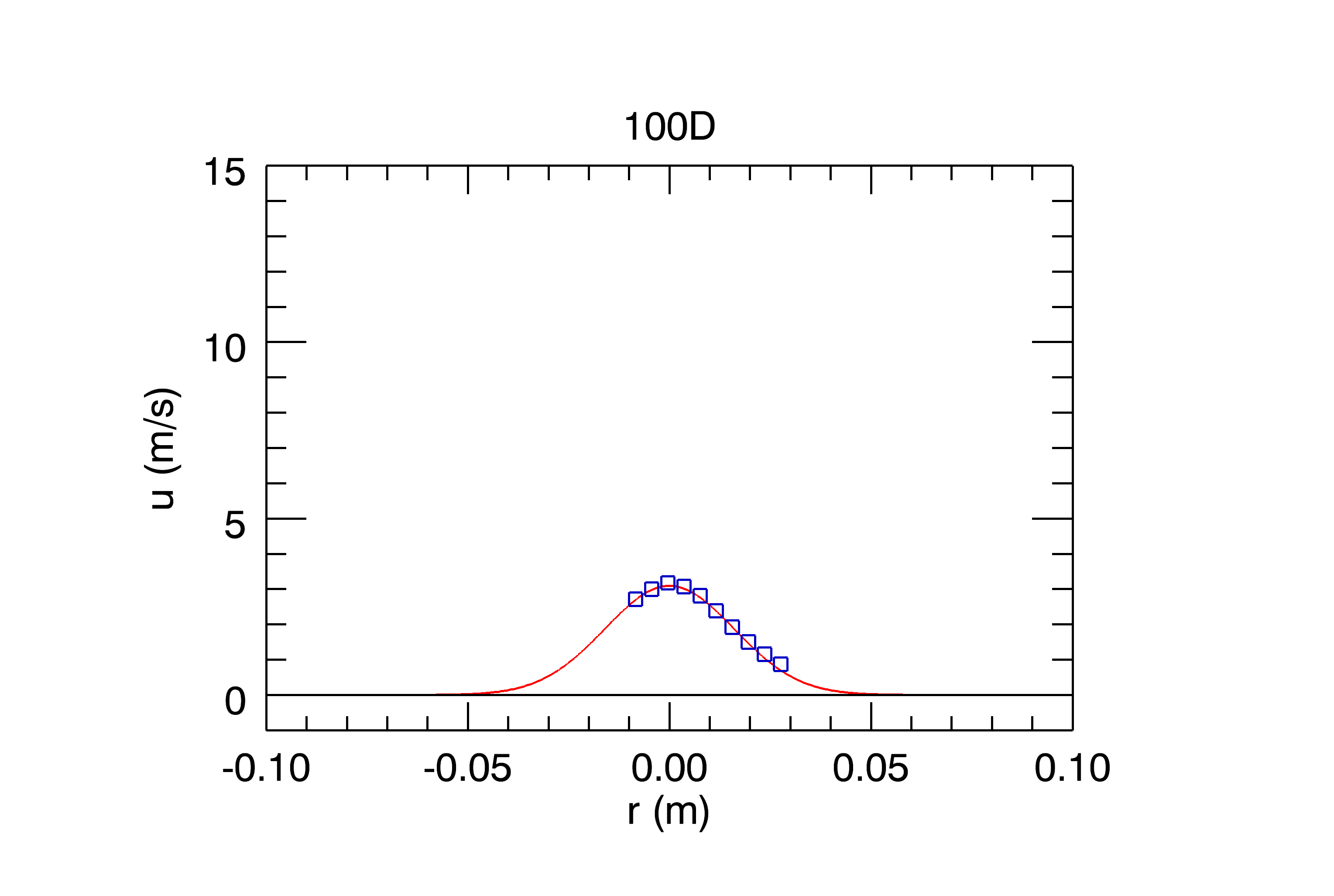}
    \end{subfigure}
    \caption{\label{fig:meas10mmjet} Measured streamwise velocity (blue squares) as a function of radial position and computed axial velocity (red curves) in the 10 mm jet.}
\end{figure*}

The optical configuration and detection circuit was the same as in the 50 mm jet measurements. This region is the self-similar region, and the measurements and instrumentation were the same as in~\cite{zhu2022similarity}. Figures~\ref{fig:meas10mmjet}a-h display the mean velocity profile at 30D to 100D in 10D steps. Each plot shows the measured velocities (blue squares, adapted from~\cite{zhu2022similarity}) overlayed with a velocity profile computed with the recursive program.

\section{Discussion}

The computed mean velocity profiles, the analytical solutions and the measurements in the 100 mm, 50 mm and 10 mm jets agree remarkably well all the way from the jet exit and through the self-similar region. Only two parameters were required to obtain the match: The initial momentum transport rates of course have to be equal for the computer models and the real jets. And 
the turbulent viscosity must be chosen in order to match the spreading angle of the computed and measured jet. However, some questions remained unanswered after the measurements, and these will be taken up in the following discussion.

\subsection{Effect of total viscosity and initial profile on far field properties}

In order to match the spreading angle of the computed jet to the measured one, the turbulent viscosity had to be adjusted. Near the jet exit, a total viscosity value close to the value for molecular viscosity, valid for a more laminar flow (about $2.5\cdot 10^{-5} \mathrm{m^2s^{-1}}$), was used, whereas a value 10 – 15 times greater was used in the far field region. This is in close agreement with discussions in the literature, where the connection between the structure of the turbulence and the spreading angle has been studied. For example,~\cite{burattini2004effect} discusses how modification of the inlet turbulence influences the statistical moments. In~\cite{xu2002effect} it is shown that a top-hat jet and a fully developed pipe flow exit profile with the same exit diameter have different spreading angles in the far field. 

We have used the recursive program to test the influence of the initial velocity profile and the magnitude of the turbulent viscosity on the spreading angle in the far field. First, we compare the far field velocity distribution for three incident profiles normalized to the same momentum flow rate: A top-hat distribution, a laminar pipe flow and a turbulent pipe flow. We use the following analytical expression for the three cases in Table~\ref{tab:table2}, where $R=D/2=0.05$ m is the jet exit radius and we have chosen $p=100$ and $n=6$.

\begin{table}[!h]
    \centering
    \caption{Incident velocity profiles: Top-hat, laminar pipe flow and turbulent pipe flow.}
    \begin{tabular}{|c|c|c|}\hline
        Top-hat & \textcolor{blue}{Laminar pipe flow} & \textcolor{red}{Turbulent pipe flow}\\\hline
        $u(r,t_0) = U_{top-hat}\exp \left ( -\frac{r}{2R} \right )^p$ & $u(r,t_0) = U_{lam.\,pipe} \left ( R^2 - r^2 \right )$ & $u(r,t_0) = U_{turb.\,pipe} \left ( 1- \left ( \frac{r}{R} \right )^2 \right )^{1/n} $\\\hline
    \end{tabular}
    \label{tab:table2}
\end{table}

\begin{figure*}[!h]
    \centering
    \begin{subfigure}{0.5\textwidth}
        \centering
        \includegraphics[width=\linewidth]{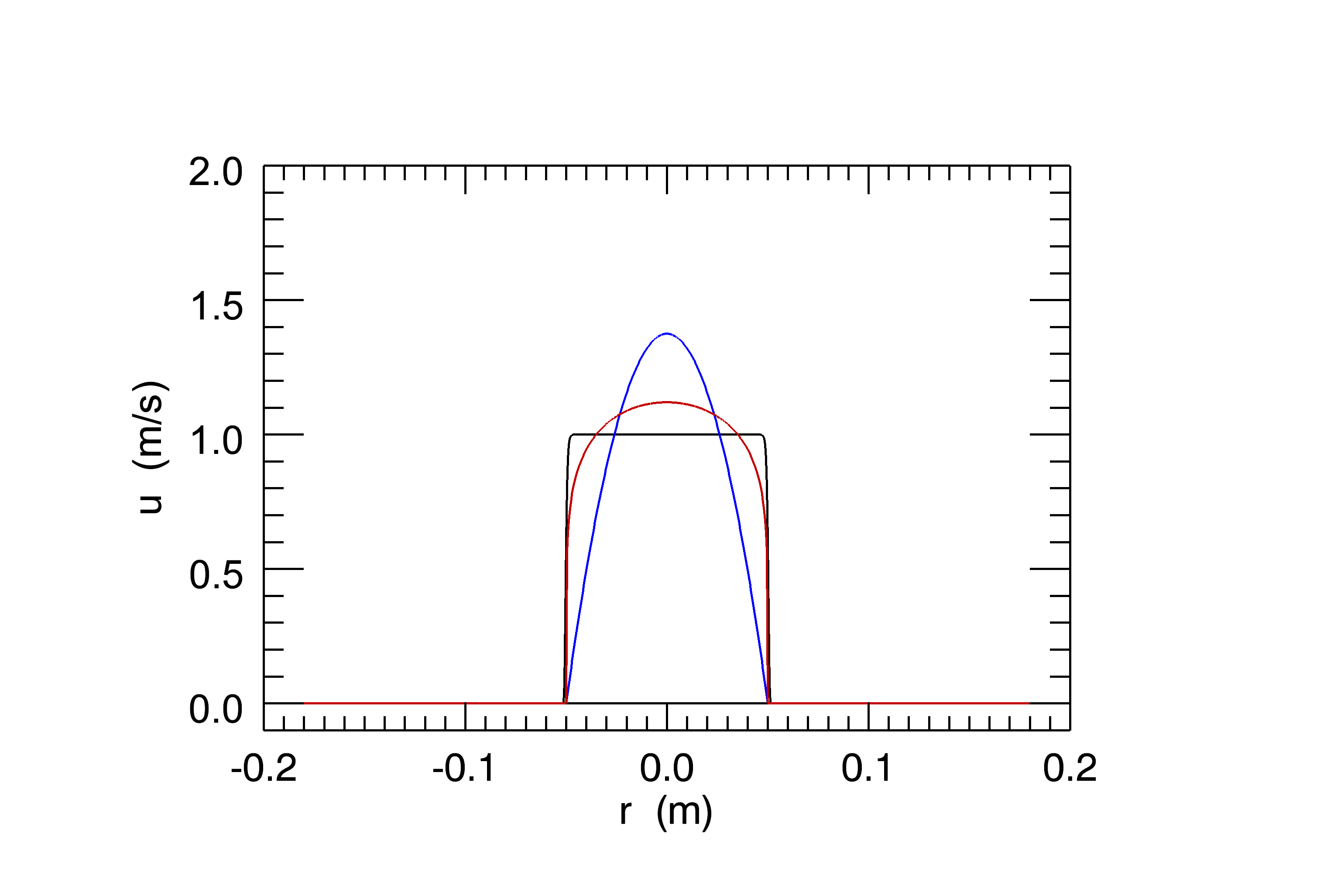}
    \end{subfigure}%
    ~ 
    \begin{subfigure}{0.5\textwidth}
        \centering
        \includegraphics[width=\linewidth]{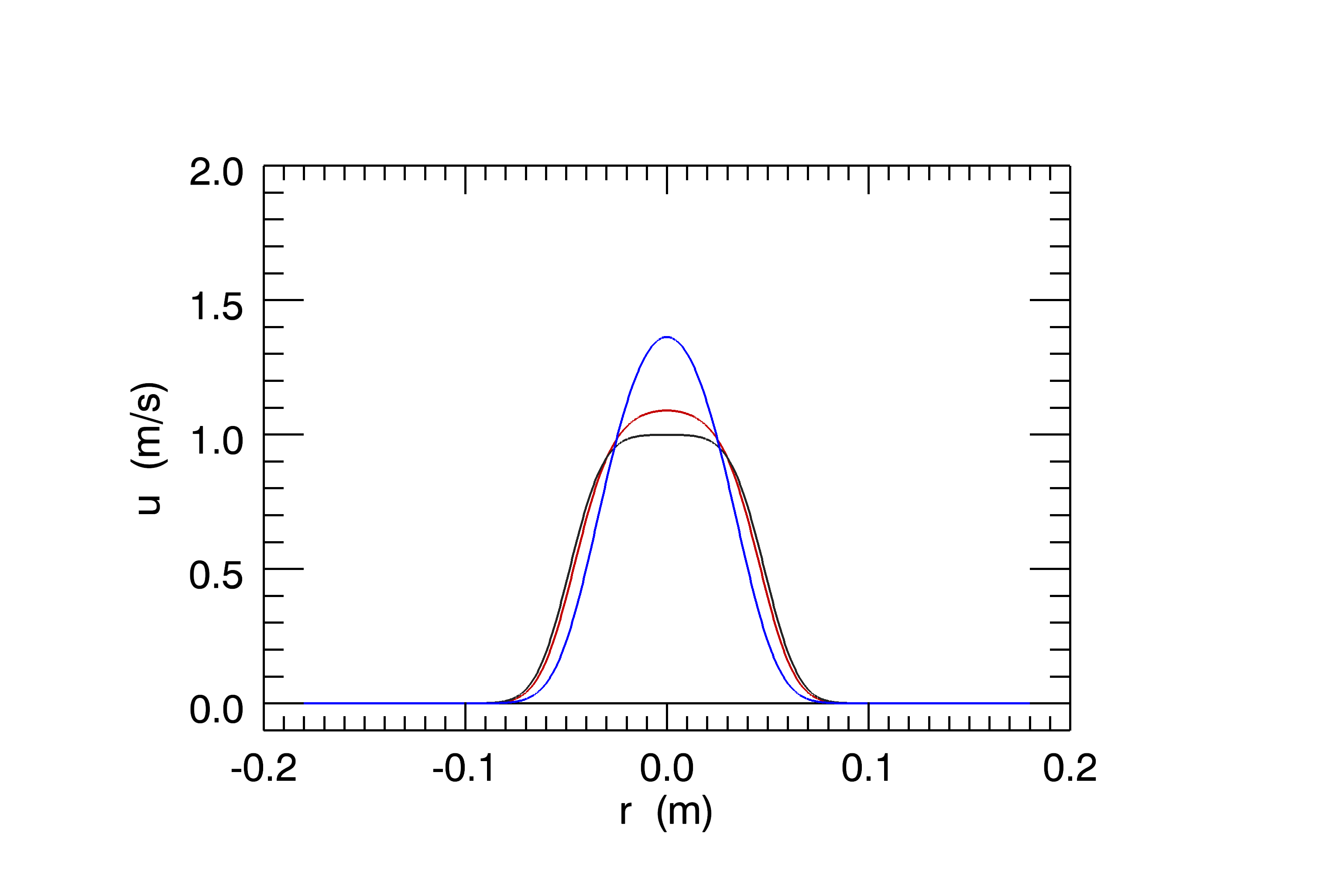}
    \end{subfigure}
        ~ 
    \begin{subfigure}{0.5\textwidth}
        \centering
        \includegraphics[width=\linewidth]{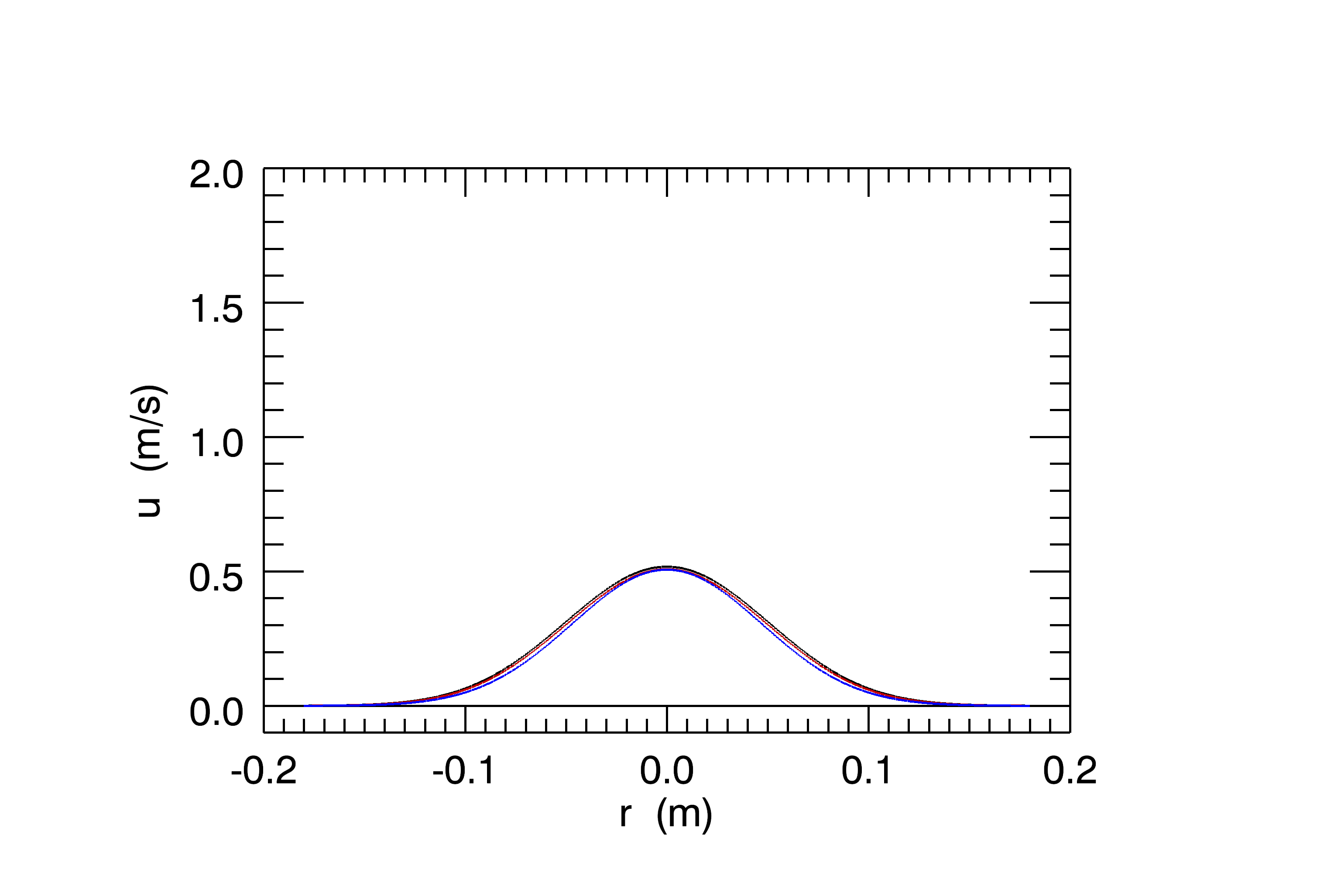}
    \end{subfigure}
    \caption{\label{fig:differentinitialprofiles} Black: Top-hat. Blue: Laminar pipe flow. Red: Turbulent pipe flow. (a) Initial velocity profiles. (b) Velocity profiles after $10\,000$ time steps. (c) Velocity profiles after $100\,000$ time steps.}
\end{figure*}

Figure~\ref{fig:differentinitialprofiles} shows the development of these three initial velocity profiles as a function of time. The different initial profiles, color coded according to Table~\ref{tab:table2}, develop into the same far field profile in the self-similar region.

\begin{figure*}[!h]
    \centering
    \begin{subfigure}{0.5\textwidth}
        \centering
        \includegraphics[width=\linewidth]{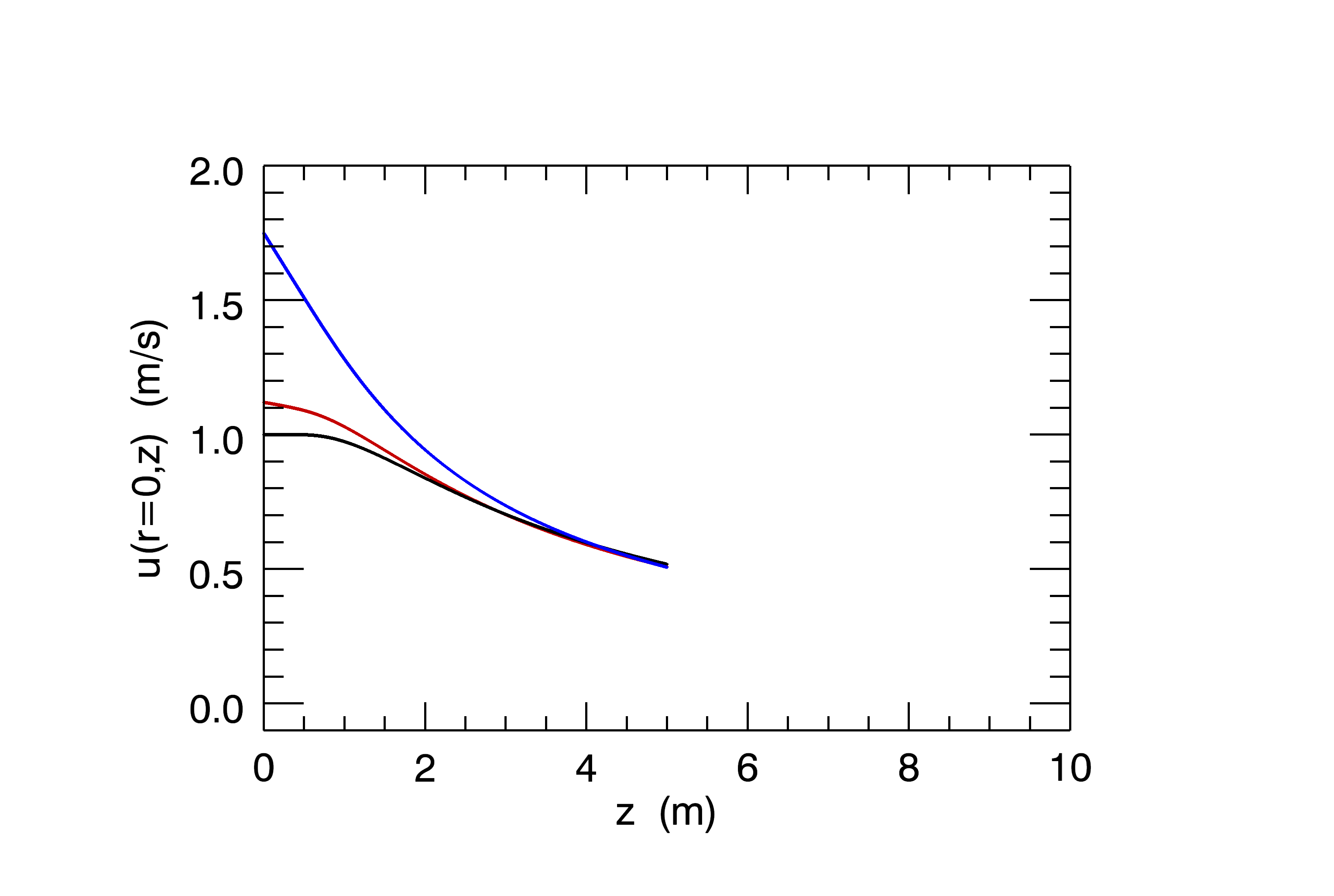}
    \end{subfigure}%
    ~ 
    \begin{subfigure}{0.5\textwidth}
        \centering
        \includegraphics[width=\linewidth]{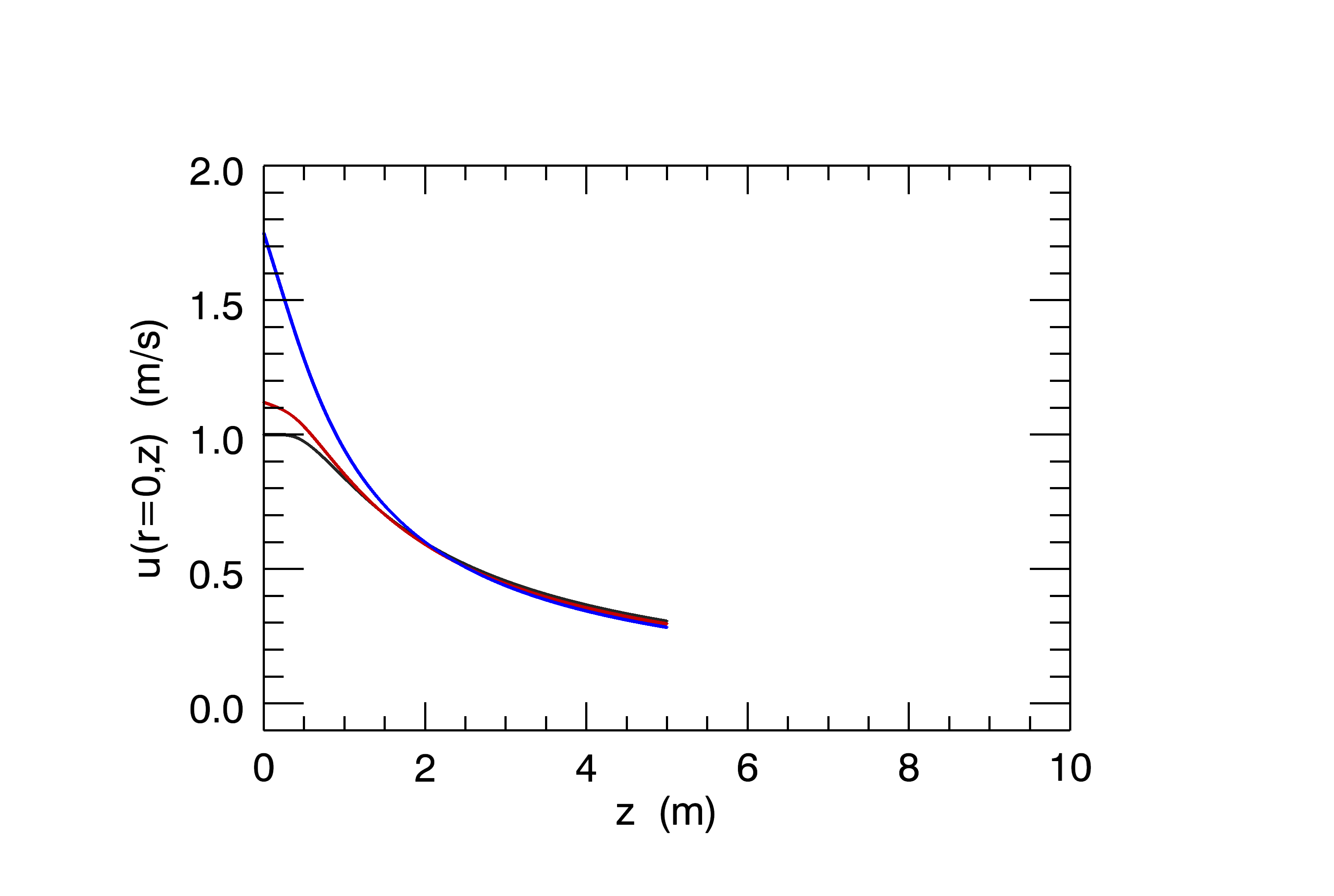}
    \end{subfigure}
    \caption{\label{fig:centerlineturbvisc} Black: Top-hat. Blue: Laminar pipe flow. Red: Turbulent pipe flow. (a) Centerline velocity from zero to $100\,000$ time steps with turbulent viscosity $\nu_T = 20 \cdot 10^{-5}$ m$^2$s$^{-1}$. (b) Corresponding plot for a turbulent viscosity of $\nu_T = 40 \cdot 10^{-5}$ m$^2$s$^{-1}$.}
\end{figure*}

In Figure~\ref{fig:centerlineturbvisc}(a), the downstream centerline velocity development is plotted for the same three initial profiles from Table~\ref{tab:table2}. For all three cases, we have used the same value for the turbulent viscosity, $\nu_T = 20$ m$^2$s$^{-1}$. Thus, it is not the shape of the initial profile that matters for the asymptotic state of the jet; they all converge to the same self-similar profile when the initial momentum transport rate is the same. What is important is the total viscosity, where the turbulent viscosity dominates in the present cases. The curves in Figure~\ref{fig:centerlineturbvisc}(b) are computed for the same three cases with the same parameters as in Figure~\ref{fig:centerlineturbvisc}(a), except with a turbulent viscosity twice as high, $\nu_T = 40$ m$^2$s$^{-1}$. The higher turbulent viscosity results in a more efficient momentum diffusion, which can be seen as expected to be a qualitatively similar, but faster, process.


\begin{figure}[!h]
\centering
\includegraphics[width=0.5\linewidth]{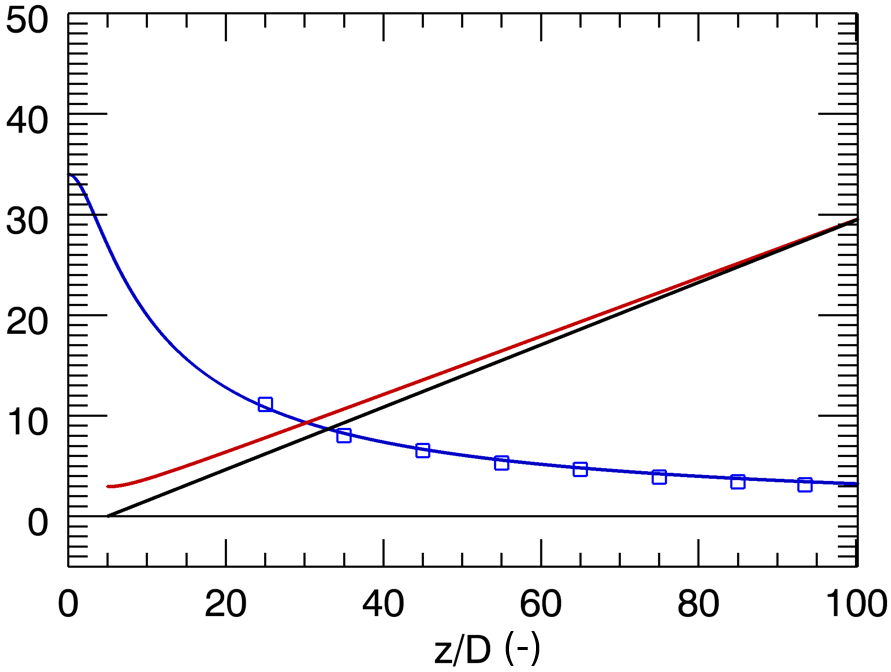}
\caption{\label{fig:CLturbvisc} Centerline average velocity of the $D= 10$ mm experimental jet with a nearly top-hat exit profile as a function of downstream distance $z/D$. Blue line computed; blue squares measured~\cite{zhu2022similarity}. Also shown is the inverse of the centerline velocity compared to a linear curve from the virtual origin at $z/D= 5$.}
\end{figure}

In Figure~\ref{fig:CLturbvisc}, we study the centerline development of the 10 mm diameter jet with a nearly top-hat exit velocity profile (super-Gaussian) all the way from the jet exit to beyond the self-similar region, computed with the recursive program and measured~\cite{zhu2022similarity}. We display both the centerline velocity and the inverse centerline velocity compared to a straight line from the virtual origin at $z/D=5$. 

\newpage

\subsection{Effect of Reynolds number on jet spreading}

\begin{figure}
\centering
\includegraphics[width=0.65\linewidth]{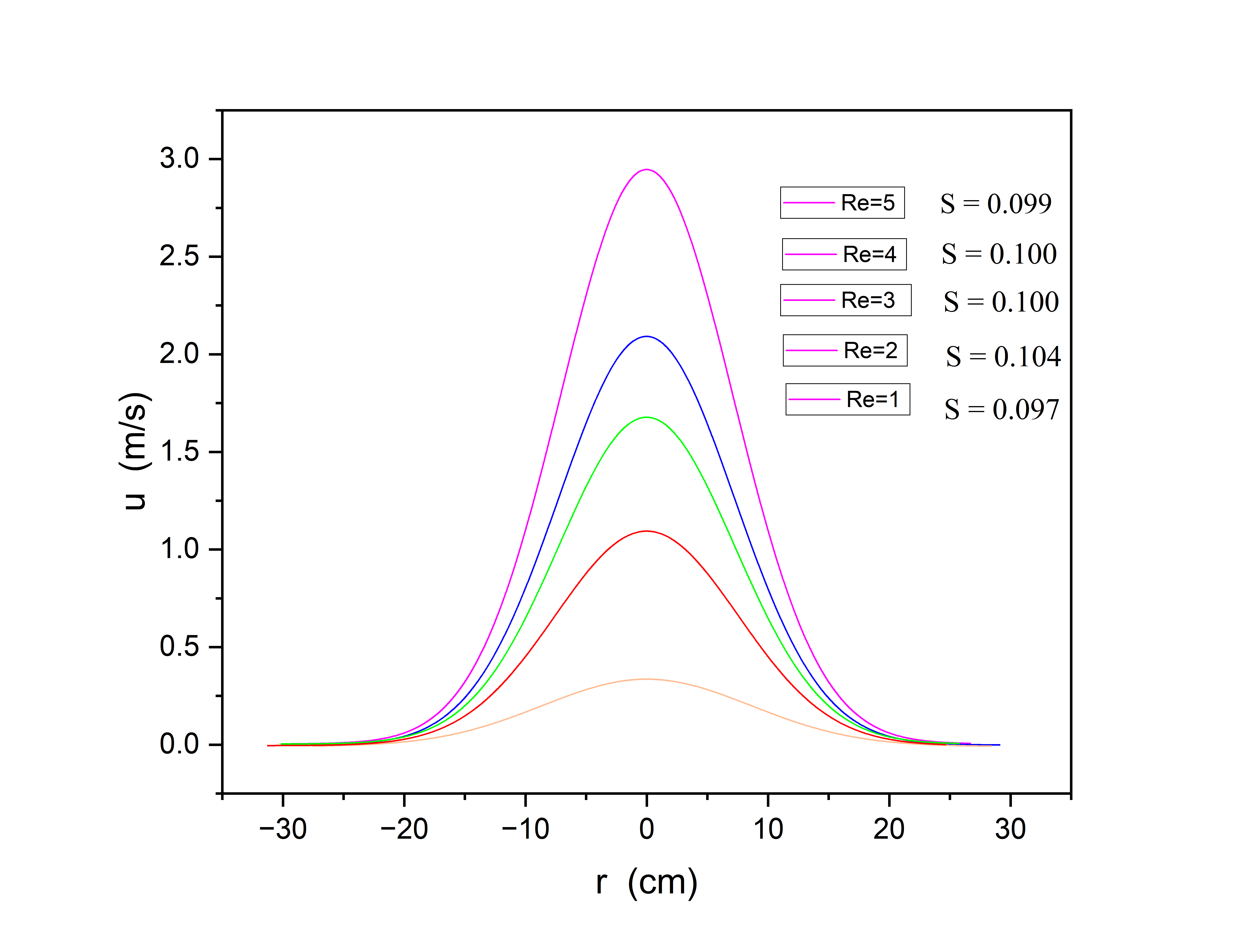}
\caption{\label{fig:Redependence} Mean velocity profiles at $z/D=90$ for different Reynolds numbers.}
\end{figure}

The Reynolds number does not enter the recursive computer program, and the Reynolds number is also not expected to have any noticeable effect on the measured cross sections. However, to ascertain this, we did a series of measurements of the spreading factor on the 10 mm diameter jet at a downstream distance of $z/D = 90$. Figure~\ref{fig:Redependence} shows the measured mean velocity profile at $z/D = 90$ for different Reynolds numbers referred to the exit velocity. The measured spreading factor $S=r_{1/2}/z_{90}$ for the five different Reynolds numbers measured on the 10 mm jet is given in Table~\ref{tab:spreadingrates}. $r_{1/2}$ is the jet half-width, i.e. where the mean velocity is half of the maximum, and $z_{90} = 90D$ is the downstream distance of the measurement position. \\

\begin{table}
    \centering
    \caption{Jet spreading factor for a range of Reynolds numbers.}
    \begin{tabular}{|c|ccccc|}\hline
        Reynolds number &  $32\,000$	& $23\,000$	& $18\,000$	& $12\,000$	& $3\,200$ \\\hline
        Spreading factor & 0.099	& 0.100	& 0.100	& 0.104	& 0.097\\\hline
    \end{tabular}
    \label{tab:spreadingrates}
\end{table}

Using the recursive formula with an initial super-Gaussian profile ($p=100$) with a 5 mm half width, we find a spreading factor equal to 0.100 within a couple of percent, confirming the universality of the jet properties and the validity of the derivation of the jet properties based entirely on Galilean symmetry properties.

\subsection{Radial velocity}

\begin{figure}
\centering
\includegraphics[width=0.65\linewidth]{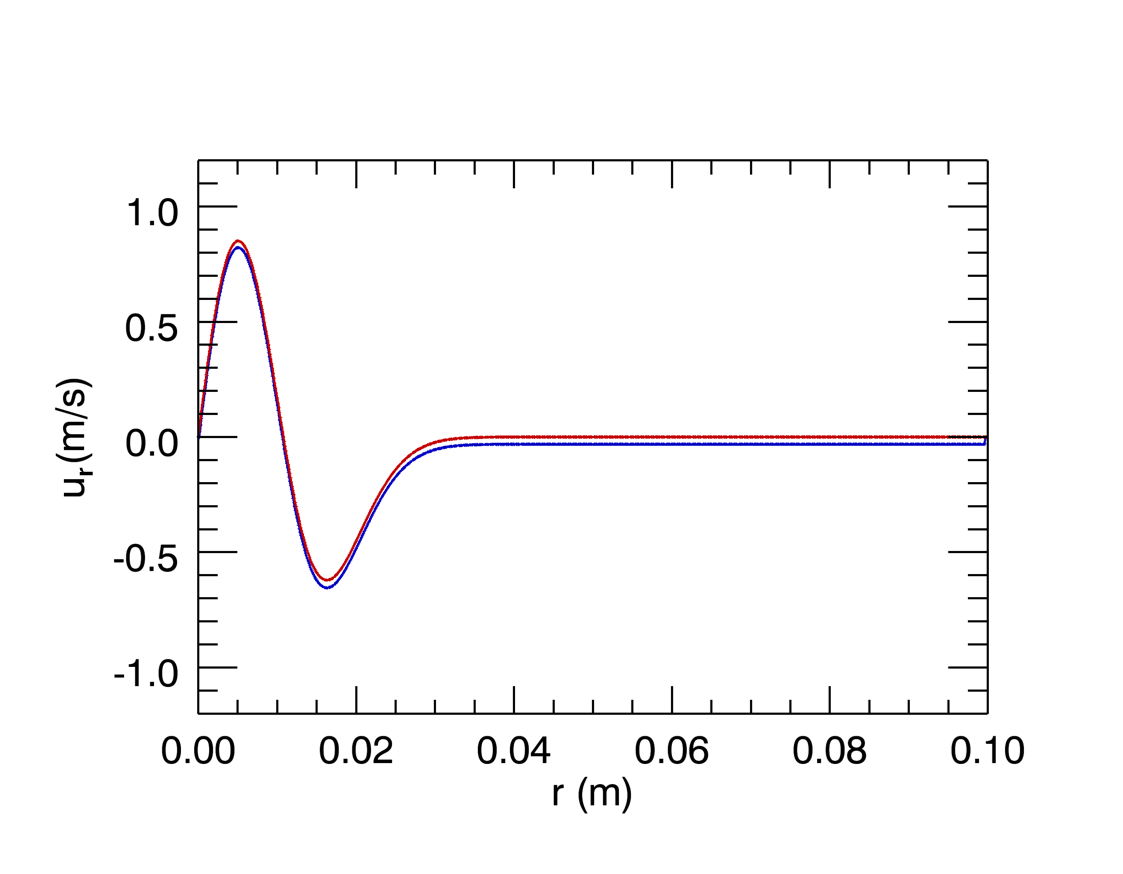}
\caption{\label{fig:radialvelocity} Radial velocity of the 10 mm jet at $z/D=100$ and $Re=6\,622$. Blue Curve: Radial velocity from broadening due to viscous and turbulent diffusion. Red curve: Radial velocity inferred from continuity.}
\end{figure}

The lateral velocity reflects the shifting and relocation of momentum as the jet develops downstream. In accordance with our model, the broadening of the jet is due to lateral diffusion of axial momentum caused by internal friction described by the turbulent kinematic viscosity. Figure~\ref{fig:radialvelocity} shows the radial velocity for a 10 mm diameter jet at $z/D=100$ with $Re=6\,622$. Integrated over the outer surface of the jet, a net influx of mass (entrainment) is evident. Often, the radial component of mean velocity is inferred from continuity. We have also used this method, and we obtain nearly identical plots, see the blue (diffusion) and red (continuity) velocity profiles overlapping in Figure~\ref{fig:radialvelocity}. Again, we emphasize that our results do not rely on self-similarity or experiment and are valid throughout the jet.

\section{Conclusion}

In this paper and in~\cite{zhu2022similarity} and~\cite{buchhave2022similarity}, we have studied the properties of the free, axisymmetric jet in air from the perspective of universality caused by basic symmetry properties of time and space, described in the Galilei Symmetry Group. In~\cite{buchhave2022similarity}, we showed that the symmetry properties mean that the jet in the region away from the jet orifice and beyond the developing region from approximately 30 jet diameters is bound to develop self-similarity, and that the self-similar region is governed by a single scaling parameter, namely the distance from the virtual origin. In~\cite{zhu2022similarity} we substantiated this by careful measurements in the self-similar region. In the present work we have focused on the developing region, in particular from the jet exit to about 10 – 15 diameters along the downstream direction. We assumed only that the jet is subject to the constraints described by the Galilei Group and that the only force acting on the fluid is Newtonian friction forces described by the turbulent dynamic viscosity. A small computer program built on these ideas and a corresponding solution to a second order partial differential equation were able to compute mean axial velocity profiles both in the region where the jet develops from the initial mean velocity profile at the jet exit and all the way through to self-similar region. The results agree exceptionally well and illustrate how the mean properties of a free, turbulent flow are formed by the fundamental symmetry properties of the surrounding world. 

Statistical functions involving the velocity of free flows are not accidental or random, but are governed by fundamental symmetries of time and space. We believe that further exploration of these phenomena will propel the understanding of developing turbulence as well as other self-similar flows and will also have important practical benefits.

\newpage

\bibliographystyle{alpha}
\bibliography{sample}

\end{document}